\documentclass[oupdraft]{sysbio}
\usepackage[T1]{fontenc}
\usepackage{lmodern}
\usepackage[dvipsnames,rgb]{xcolor}
\usepackage{times}

\usepackage[colorlinks=true, allcolors=blue]{hyperref}
\usepackage{cleveref}
\usepackage{nameref}

\usepackage[export]{adjustbox}
\usepackage{graphicx}
\usepackage[caption=false]{subfig}
\usepackage{array}
\usepackage{booktabs}

\usepackage{url}
\usepackage{tikz}
\usetikzlibrary{decorations.pathmorphing, decorations.pathreplacing, positioning,shapes.geometric}
\tikzset{every picture/.style={/utils/exec={\sffamily}}}

\usepackage{algorithm}
\usepackage{algpseudocode}

\usepackage{dsfont}

\DeclareRobustCommand{\sbseries}{\fontseries{sb}\selectfont}
\DeclareTextFontCommand{\textsb}{\sbseries}

\begin{document}

\title{Generalised Bayesian distance-based phylogenetics for the genomics era}

\author{Matthew J Penn$^{1\ast}$, Neil Scheidwasser$^{2\ast}$, Mark P Khurana$^{2}$, Christl A Donnelly$^{1, 3}$, David A Duch\^ene$^{2\ast}$, and Samir Bhatt$^{2, 4\ast}$\\[4pt]
\textit{$^{1}$Department of Statistics, University of Oxford, Oxford, United Kingdom}
\\
\textit{$^{2}$Section of Epidemiology, University of Copenhagen, Copenhagen, Denmark}
\\
\textit{$^{3}$Pandemic Sciences Institute, University of Oxford, Oxford, United Kingdom}
\\
\textit{$^{4}$MRC Centre for Global Infectious Disease Analysis, Imperial College London, London, United Kingdom}
\\
\textit{$^{\ast}$Equal contribution}
\\[2pt]
Correspondence: \textit{david.duchene@sund.ku.dk}, \textit{samir.bhatt@sund.ku.dk}}

\markboth%
{Penn M., Scheidwasser N., et al.}
{A phylogenetic likelihood from BME}

\maketitle

\begin{abstract}
{As whole genomes become widely available, maximum likelihood and Bayesian phylogenetic methods are demonstrating their limits in meeting the escalating computational demands. Conversely, distance-based phylogenetic methods are efficient, but are rarely favoured due to their inferior performance. Here, we extend distance-based phylogenetics using an entropy-based likelihood of the evolution among pairs of taxa, allowing for fast Bayesian inference in genome-scale datasets. We provide evidence of a close link between the inference criteria used in distance methods and Felsenstein's likelihood, such that the methods are expected to have comparable performance in practice. Using the entropic likelihood, we perform Bayesian inference on three phylogenetic benchmark datasets and find that estimates closely correspond with previous inferences. We also apply this rapid inference approach to a 60-million-site alignment from 363 avian taxa, covering most avian families. The method has outstanding performance and reveals substantial uncertainty in the avian diversification events immediately after the K-Pg transition event. The entropic likelihood allows for efficient Bayesian phylogenetic inference, accommodating the analysis demands of the genomic era.}
{phylogenomics, Bayesian inference, distance phylogenetics, avian phylogeny}
\end{abstract}

\section{Introduction}

The field of phylogenetics underpins a large portion of modern biological research, offering a powerful framework to describe branching processes across the tree of life. With the advent of vast amounts of genomic data, the field is now challenged by the fact that the space of possible phylogenetic trees scales double-factorially: for $n$ samples (or leaves) there are $1 \cdot 3 \cdot 5 \ldots (2n-3)$ number of possible bifurcating unrooted trees~\citep{Cavalli-Sforza1967-xh, Felsenstein1978-qc}. This makes it impossible to consider every possible tree as a candidate, except in very small datasets (approximately $\leq$ 10 taxa). Whilst phylogenetic reconstruction under any existing method is NP-hard~\citep{roch2006}, methods based on Felsenstein's likelihood~\citep{felsenstein1981} are generally considered highly robust~\citep{yang2012}, where various heuristic approaches to tree search have shown excellent performance in both simulations~\citep{stamatakis2014,nguyen2015} and in comparisons with independent data from the fossil record~\citep{Morlon2011-pe}. This has led maximum likelihood to become the underlying criterion for many phylogenetic tree estimation frameworks~\citep{stamatakis2014, minh2020, demaio2023}. However, even under highly efficient software architectures, the computational demand of maximum likelihood approaches is becoming impractical for many genome-scale and epidemiological datasets, raising questions about the feasibility of these analyses into the future of the genomics era~\citep{kumar2022, simion2020}.

The primary caveat with modern phylogenetics is that the traditional Felsenstein's likelihood~\citep{felsenstein1981} is highly complex to compute for large datasets. Specifically, calculating the likelihood of a \emph{single} tree is $\mathcal{O}(nNc^2)$ for $n$ leaves, $N$ unique site patterns and $c$ character states. Even with the efficiency savings that occur when considering similar trees, the calculation remains highly complex, being of at least $\mathcal{O}(N)$. Thus, optimising by making the best move on the tree topology using subtree-pruning-and-regrafting (SPR) from the set of $n^2$ possible moves still gives a high complexity, of at least $\mathcal{O}(n^2 N k_L)$ for $k_L$ optimisation steps. Conversely, distance-based approaches using the balanced minimum evolution (BME) criterion~\citep{pauplin2000,Desper2002-uy} have a high initial pre-processing complexity, of $\mathcal{O}(Nn^2 + c^2n^2)$, but far lower optimisation complexity that is independent of the number of unique site patterns $N$. Indeed, using a similar SPR scheme under this criterion has a worst-case complexity of $\mathcal{O}(k_Bn^2 $\text{Diam}(T)$)$ for $k_B$ optimisation steps, where $\text{Diam}(T) \lesssim \mathcal{O}(n)$ is the maximum number of edges between each pair of taxa. In sum, the statistical benefits of using Felsenstein's likelihood can be outweighed by their high computational cost, providing a strong motivation for developing alternative fast approaches for phylogenetic inference.

The gap in complexity between likelihood and distance methods widens dramatically in Bayesian inference settings, where large numbers of likelihoods are evaluated. When analysing whole genomes with millions of sites, or thousands of taxa, likelihood methods become prohibitively slow and the only alternatives are to perform extensive data filtering, or to use a fast approach like minimum evolution or maximum parsimony. While maximum parsimony is consistent under certain conditions, BME has been proven to be more broadly statistically consistent under knowledge of the true model and in data with no recombination~\citep{desper2004}. Although not as empirically accurate as maximum likelihood, the statistical guarantees of BME will ensure it is highly accurate in large-scale data regimes, particularly when considering branch supports and other forms of uncertainty (e.g., in branch lengths or date estimates)~\citep{lefort2015}.

Although a point estimate of a single best tree can be useful, uncertainty quantification is another critical aspect of biological hypothesis testing, yet generally prohibitive in very large modern datasets. Existing solutions include bootstrapping~\citep{felsenstein1985, Efron1996-gv} and Bayesian posterior supports~\citep{Rannala1996-zk}. For all inferential approaches, bootstrapping is possible and has highly efficient implementations~\citep{Hoang2018-mi}. However, the interpretation of a bootstrap sample is non-probabilistic. It is a measure of re-sampling variance and of the robustness of the estimator over small changes in the data~\citep{Holmes2003-cg}. In contrast, Bayesian approaches yield the posterior probability for any given tree and for branches within a tree. Bayesian approaches, however, require a likelihood and are therefore unsuitable for inference under minimum evolution and maximum parsimony. Generalised Bayesian approaches~\citep{Bissiri2016-nl} offer an alternative, but are as of yet underdeveloped for phylogenetics. Therefore, a key limitation of minimum evolution is that the objective criterion is not a likelihood and therefore cannot be used for hypothesis testing or, more importantly, for Bayesian analysis. Until now, uncertainty under this framework has only been quantified via bootstrapping.

To address the various shortcomings of phylogenetic methods for inference using large modern datasets, we develop a new likelihood function by considering the effect of using tree entropy as a prior distribution. Using simplifying approximations, we reduce this new likelihood to a tree length given an ``entropic distance matrix'' which we denote as $d^S_{ij}$. This entropic distance is the expected entropy of the evolution process across the path between a pair of taxa in the tree, given their genetic distance. The resulting entropic likelihood is highly correlated with the well-established likelihood of Felsenstein~\citep{felsenstein1983} that underpins all maximum likelihood and Bayesian phylogenetics (Fig.~\ref{fig:entropic_vs_fels}), and we provide a mathematical justification for the closeness of this relationship. Through studying the similarities between the entropic distance $d^S_{ij}$ and standard genetic distance $d_{ij}$, we find that, under a Markovian branching process, the entropic distance is extremely well-approximated by a linear function of the standard genetic distance (Fig.~\ref{fig:entropy}). Specifically, given a continuous time Markov chain (CTMC) substitution model $P$ (e.g., Jukes-Cantor model~\citep{jukes1969}), the entropy along a single branch is $S(t) = \sum_{a,b}\pi_aP_{ab}(t)\log(P_{ab}(t))$ (where $\pi$ denotes stationary frequencies). Furthermore, the entropic tree length is approximately a linear function of the BME tree length. The error of this approximation has favourable mathematical properties (Fig.~\ref{fig:error}), and this new formulation provides additional insights into the robustness of BME as an optimality criterion for phylogenetic inference. Our work paves the way for Bayesian phylogenetic inference using distance-based methods, allowing for highly efficient probabilistic analysis of the massive datasets that define the genomic era.

\bigskip
\section{Materials and Methods}
\subsection{Notation and Preliminaries}
Throughout this paper, we will use $\mathcal{T}$ to denote the true topology which, using the phylogenetic data, we are aiming to infer (see summary of terms in Table~\ref{tab:params}). $\mathcal{U}$ will be used for the tree topology under consideration - which may or may not be the true tree. Note that our methods are primarily designed to work with ultrametric trees, though could be extended to non-ultrametric models. 

The derivation of the phylogenetic likelihood has been well established in the literature \cite{felsenstein1983}. However, we will revisit this derivation to establish consistent notation throughout the paper and to explicitly highlight the mathematical connections between balanced minimum evolution and the phylogenetic likelihood.

\begin{table*}[h!]
\centering
\caption{Common parameter definitions in this paper and the appendix}
\begin{tabular}{llrrll}
\toprule
\textbf{Parameter} & \textbf{Definition} \\
\toprule
$\boldsymbol{b}$ & Branch lengths for a given phylogeny \\
$d_{ij}$ & Evolutionary distance between taxa $i$ and $j$ \\
$d^S_{ij}$ & Entropic distance between taxa $i$ and $j$ \\
$H$ & Entropy across some or all nodes in the phylogeny \\
$K$ & Expected instantaneous substitution rate \\
$\ell_F$ & Felsenstein's likelihood \\
$\ell_S$ & Entropic likelihood \\
$p(g,\mathcal{U},\boldsymbol{b})$ & Probability of observing a site pattern $g$ on a phylogeny (internal and \\
 & external nodes) with topology $\mathcal{U}$ and branch lengths $\boldsymbol{b}$ \\
$P_{ij}(t)$ & Probability of a site with state $i$ at time 0 being state $j$ at time $t$ \\
$\pi_i$ & Stationary distribution probability of a site having value $i$ \\
$Q$ & Transitions matrix under CTMC \\
$S(t)$ & Entropy of an observation of a site pattern after time $t$ \\
$\theta$ & Branching rate of our phylogenetic construction model \\
$\mathcal{T}$ & True topology of the tree under consideration \\
\bottomrule
\end{tabular}
\label{tab:params}
\end{table*}

The classical maximum likelihood problem in phylogenetics involves the construction of a weighted, binary tree with topology $\mathcal{U}$ and branch lengths $\boldsymbol{b}$, describing the evolutionary history of a set of $n$ taxa, each of which corresponds to a leaf node of $\mathcal{U}$.

In general, a fixed number of sites of genomic or proteomic data is available for each taxon. Typically, the substitution process of a single site is modelled using a reversible continuous time Markov chain (CTMC) where the transition matrix/kernel is:
\begin{equation}
\label{eq:Pijeqn}
    P_{ij}(t; \boldsymbol{\phi}) = \mathbb{P}(\text{A site changes from state $i$ to state $j$ in time $t$})
\end{equation}
where $\boldsymbol{\phi}$ represents the set of model parameters, which are generally estimated as part of the inference process. Estimating these parameters is a standard procedure, and so we will not refer to $\boldsymbol{\phi}$ in the methods. Note, we also do not attempt to jointly estimate $\boldsymbol{\phi}$ from the data alongside the tree.

We use $Q$ to denote the Q-matrix of the substitution CTMC, and define $\pi$ to be its stationary distribution. We assume that $|Q_{ij}| > 0$ (and hence $P_{ij}(t) > 0$) for all $t > 0$ and all $i$ and $j$. This is not strictly necessary for the results presented in this paper but simplifies the derivations.

To simulate the sites of a tree (and to calculate likelihoods), one must choose a node, $r$, denoted hereinafter as the \textit{simulation root} - which corresponds to the node at which to begin the simulation. It will be assumed that the sites at node $r$ are chosen according to the stationary distribution of the CTMC. It is shown in Lemma~\ref{lem:sim} in the Appendix that the distribution of the sites of the tree is independent of which node is chosen to be the simulation root.

Thus, given a set of characters $g$ for a given site on the internal \emph{and} external nodes (totalling 2n-1), one can define a function $p(g,\mathcal{U},\boldsymbol{b})$ which gives the likelihood of these values being observed for a given site on the topology $\mathcal{U}$ with branch lengths $\boldsymbol{b}$. This can be done by considering each edge separately, and will be derived later in the methods.

\subsubsection{Felsenstein's likelihood}
In general, only the characters on leaf nodes $l(g)$ are observed, rather than the full set of characters $g$.

Felsenstein's likelihood solves this problem by marginalising over the possible values of the sites of the internal nodes. For a site with characters $l(g)$ on the leaf nodes, the likelihood of that site (on the topology $\mathcal{U}$ with branch lengths $\boldsymbol{b}$) is then
\begin{equation}
\label{eq:FelForm}
    \lambda_F(l(g),\mathcal{U},\boldsymbol{b}) = \sum_{h\in \mathcal{G} : l(h) = l(g)} p(h,\mathcal{U},\boldsymbol{b})
\end{equation}
where $\mathcal{G}$ denotes the set of possible site patterns over the whole tree.

Calculating the likelihood directly from (\ref{eq:FelForm}) is inefficient, and in practice, $\lambda_F$ can be computed through a postorder tree traversal, where one iteratively finds the likelihood of the tree rooted at each node $x$, conditional on the value of the sites at $x$.

Under the assumption that different sites evolve independently, Felsenstein's likelihood $\ell_F$ of the tree is then simply the product of the $\lambda_F(g,\mathcal{U},\boldsymbol{b})$ for each observed set of characters $g$.

\subsection{Motivation: A likelihood from balanced minimum evolution}
In the balanced minimum evolution paradigm~\citep{pauplin2000}, the aim is to find the tree where, under a specific (balanced) method of branch length estimation, the length of this tree (that is, the sum of its branch lengths) is minimised. To produce a similarly-motivated likelihood for each possible tree (comprised of a topology $\mathcal{U}$ and branch lengths $\boldsymbol{b}$), we develop a measure, $E(\mathcal{U},\boldsymbol{b})$, of the entropy of this tree, defined explicitly in the subsequent section. This provides a similar measure of tree complexity - low expected likelihood (just like high length) means that a more complex evolutionary process would have occurred.

To create this entropy measure, we suppose that we observe the sites at both internal and external nodes of the tree. If, for a random simulated mutation process on $\mathcal{U}$, we observe sites $G$ across all these nodes, then our entropy measure is simply the entropy (that is, the expected log-likelihood) of the random variable $G$.

To apply such principles to a likelihood setting, we consider a Bayesian approach to inferring the true tree. Suppose that for a tree with topology $\mathcal{U}$ and branch lengths $\boldsymbol{b}$, our prior distribution is $\psi(\mathcal{U},\boldsymbol{b})$. Suppose that the likelihood of observing the data, $\mathcal{D}$ is $P( \mathcal{D}|\mathcal{U},\boldsymbol{b})$. Then, the posterior likelihood of a topology $\mathcal{U}$ is proportional to
\begin{equation}
\label{eq:the_integral}
    L(\mathcal{U}) = \int_{\boldsymbol{B}} P(\mathcal{D}|\mathcal{U},\boldsymbol{b})\psi(\mathcal{U},\boldsymbol{b}) d\boldsymbol{b}
\end{equation}
where $\boldsymbol{B}$ denotes the set of possible values of $\boldsymbol{b}$ and $d \boldsymbol{b} = db_1db_2...db_{2n-1}$.

While the prior distribution is not explicitly considered in the classical Felsenstein's likelihood inference problem, many authors assign a prior to the branch lengths \citep{rannala2012tail}, which, in this setup, is akin to choosing $\psi$ to be a function of $\boldsymbol{b}$ (for example, using the assumption that branch lengths are exponentially- or gamma-distributed) and independent of the topology, $\mathcal{U}$.

Under our minimal-complexity paradigm, we use the tree entropy as a prior as we expect simpler trees to be more likely. As entropy has the same range as a log-likelihood (rather than a likelihood), we set (ignoring the normalisation constant),
\begin{equation}
    \log(\psi(\mathcal{U},\boldsymbol{b})) =E(\mathcal{U},\boldsymbol{b})
\end{equation}
where, as previously mentioned, $E$ is the entropy of the tree. To simplify the calculation, we make the following assumptions. Firstly, we assume that we have enough data such that (using the consistency of the BME estimators with no model misspecification \citep{desper2004}), the likelihood $P$ is closely concentrated around the BME branch lengths, $\boldsymbol{b}^*(\mathcal{D},\mathcal{U})$. That is, we use the approximation
\begin{equation}
    P(\mathcal{D}|\mathcal{U},\boldsymbol{b}) \approx  C_1 P(\mathcal{D}|\mathcal{U})\delta(\boldsymbol{b}-\boldsymbol{b}^*(\mathcal{D},\mathcal{U}))
\end{equation}
for some normalisation constant $C_1$. Secondly, we choose to ignore the additional information provided about the topology, $P(\mathcal{D}|\mathcal{U})$. Again, this mirrors the classical BME setup where the objective can be calculated from an unordered set of branch lengths. This reduces the total information used and, we conjecture, is the main reason why Felsenstein's likelihood generally outperforms balanced minimum evolution - even though both are asymptotically consistent, Felsenstein's likelihood utilises all the information from the data and will therefore perform better on finite datasets. However, the relationship between the topological posterior and the data is far more complex than the relationship between the branch length posterior and the data, so restricting ourselves to information about the branch lengths substantially reduces the complexity of the tree inference problem. Thus, ignoring normalisation, we suppose that
\begin{equation}
    P(\mathcal{D}|\mathcal{U}) = C_2
\end{equation}
for some constant $C_2$. Under these assumptions, we can evaluate the integral (\ref{eq:the_integral}) to see that $L$ becomes
\begin{equation}
    L(\mathcal{U}) = \psi(\mathcal{U},\boldsymbol{b}^*(\mathcal{U},\mathcal{D}))
\end{equation}
and hence
\begin{equation}
\label{eq:Lu}
    \ell(\mathcal{U}) = E(\mathcal{U},\boldsymbol{b}^*(\mathcal{U},\mathcal{D}))
\end{equation}
which is the expected likelihood of the tree with topology $\mathcal{U}$ and branch lengths given by the BME estimates.

In the remainder of this paper, we will derive an efficient way of calculating $L(\mathcal{U})$ and show that is closely approximated by a linear function of the standard BME objective. We will also show that this likelihood is closely related to Felsenstein's likelihood, and can therefore be used to perform approximate inference in that paradigm. 
\subsection{Derivation: Defining the entropic likelihood}
\subsubsection{Calculating \texorpdfstring{$p(g,\mathcal{U},\boldsymbol{b})$}{p(g, U, b)}}

To begin, it is helpful to find an explicit formula for the function $p(g,\mathcal{U},\boldsymbol{b})$. Define the set of edges to be $\mathcal{E} = \{(e_i^1,e_i^2)~|~i =0,1,..\}$, where $e_i^1$ and $e_i^2$ are the nodes which this edge connects, such that $e^1_i$ is the closest to the simulation root $r$. Note that the same node may be represented by different $e_j^k$ in this set definition. Suppose that $z_i^j$ is the site value on node $e_i^j$ and that $b_i$ is the length of edge $(e_i^1,e_i^2)$. Then, as the substitution CTMC on each edge is conditionally independent given the start and end values of the sites, we have  
\begin{equation}
\label{eq:pgt}
    p(g,\mathcal{U},\boldsymbol{b})= \pi_{z_r} \prod_{i}P_{z_i^1,z_i^2}(b_i)
\end{equation}
Here $P_{z_i^1,z_i^2}(b_i)$ is the probability of $z_i^1$ mutating into $z_i^2$ in time $b_i$, while $z_r$ gives the value of the site at the simulation root. Note that Lemma~\ref{lem:sim} shows that the value of $p(g,\mathcal{T})$ is independent of $r$.

\subsubsection{Calculating \texorpdfstring{$E$}{E}}
Recalling that the site pattern across all nodes is defined to be $G$, we can now note that, as 
\begin{equation}
\mathbb{E}\bigg(\log(p(G,\mathcal{U},\boldsymbol{b}))\bigg) = \mathbb{E}\bigg(\log(\pi_{Z_r}) +  \sum_{i}\log(P_{Z_i^1,Z_i^2}(b_i))\bigg) = \mathbb{E}\bigg(\log(\pi_{Z_r})\bigg) + \sum_{i}\mathbb{E}\bigg(\log(P_{Z_i^1,Z_i^2}(b_i))\bigg)
\end{equation}
We can ignore $\mathbb{E}(\log(\pi_{Z_r}))$, as it is independent of the topology and the data (this is simply the entropy of the starting genome at the simulation root). Moreover, using Lemma~\ref{lem:pairs}, we know that as the length of the branch joining $Z_i^1$ and $Z_i^2$ is $b_i$,
\begin{equation}
    (Z_i^1,Z_i^2) \stackrel{d}{=} (M_0,M_{b_i})
\end{equation}
where $M$ is an independent copy of the mutation CTMC, and $\stackrel{d}{=}$ denotes equality in distribution. Thus, we have
\begin{equation}
    \mathbb{E}\bigg(\log(P_{z_i^1,z_i^2}(b_i))\bigg) = -S(b_i)
\end{equation}
where $S$ is the entropy of our mutation CTMC run for time $b_i$ (assumed to be positive - hence the minus sign as log-likelihoods are always negative). Thus,
\begin{equation}
\label{eq:entexact}
\ell(\mathcal{U})=     \mathbb{E}\bigg(\log(p(G,\mathcal{U}))\bigg) = \text{const.} - \sum_{i}S(b_i)
\end{equation}
 For trees with $N$ sites, we can use the independence of different sites, meaning that the entropy $E$ is simply $N$ multiplied by $\mathbb{E}\big(\log(p(G,\mathcal{U}))\big)$.
\subsubsection{The entropic likelihood}
 While Equation \ref{eq:entexact} is computable, using the individual balanced estimates to the branch lengths $b_i$ can lead to problems. For example, there is no guarantee that these estimates will be positive. Having the ability to use some prior that controls branch lengths is therefore helpful and, while we cannot efficiently do this on an individual-branch level, we develop a method in this section for imposing priors on an inter-taxa scale.

To begin, note that the objective function (Equation \ref{eq:Lu}) is -1 multiplied by the length of a tree where the branches have lengths $S(b_i)$ (rather than their original $b_i$ lengths, which were the BME approximations). Thus, we seek to find an entropic distance matrix $d^S$ such that $d^S_{ij}$ is the distance between taxa $i$ and $j$ in our entropy-weighted tree (for a single site). In this case, the objective function is the length of the BME tree with distance matrix $d^S$, meaning
\begin{equation}
\label{eq:bmeobj}
    \ell_S(\mathcal{U}) = -N\sum_{i\neq j}2^{-e_{ij}} d^S_{ij}
\end{equation}
where $e_{ij}$ is the path length between $i$ and $j$ in $\mathcal{U}$. The $2^{-e_{ij}}$ terms come from the BME objective, noting that the length of a tree with inter-taxa distance matrix $D_{ij}$ is
\begin{equation}
    \sum_{i,j}D_{ij}2^{-e_{ij}}
\end{equation}
This sum over $i\neq j$ considers all ordered pairs $i$ and $j$. We call the likelihood $\ell_S$ the \textit{entropic likelihood}.
\subsubsection{Models for approximating \texorpdfstring{$d^S_{ij}$}{d\^S\_{ij}}}
If one supposes that the tree was generated according to some random process, then one can set
\begin{equation}
    d^S_{ij} =\mathbb{E}\big(D^S_{ij}(d_{ij})\big)
\end{equation}
where $D^S_{ij}(\tau)$ is a random variable with distribution equal to the entropic distance between two taxa on a tree generated according to this random process conditional on the distance between these taxa being equal to $d_{ij}$.

In this paper, we use a simple branching process where each branch has a length distributed according to some probability density function $f$, splits into two branches at the end of this length, and the process is stopped after some fixed time $T$ (so that any branches still ``active'' at time $T$ form a leaf node). We hope to examine more complicated models in the future, particularly those which allow for non-ultrametric trees (for example, by varying the mutation rate along the branches). However, as we will show in the subsequent sections, the approximate linearity of our entropic distance means that this model is still useful even in the absence of ultrametricity.

Lemma~\ref{lem:branchgen} and Lemma~\ref{lem:branchexp} provide formulae for the tree being generated by this branching process. In Lemma~\ref{lem:branchgen}, one has
\begin{equation}
    \mathbb{E}(D^S_{ij}(d_{ij})) = 2\mathcal{L}^{-1}\Bigg(\frac{\bar{S}(p) + (\frac{1}{p}-1)\bar{g}(p)}{1 - \bar{f}(p) }\Bigg)\bigg|_{\frac{d_{ij}}{2}}
\end{equation}
where $\mathcal{L}^{-1}$ is the inverse Laplace transform, overbars denote Laplace transforms and $\bar{g}(p)$ is the Laplace transform of the function $S(t)f(t)$. This explicit solution arises from the renewal equation
\begin{align}
\label{eq:renewal_eq}
  \frac{1}{2}\mathbb{E}(D^S_{ij}(d_{ij})) := h(\tau)  = (1-F(\tau))S(\tau) + \int_0^{\tau}(h(\tau - t) + S(t))f(t)dt
\end{align}
For any parametric choice of branch length distribution, these renewal equations can be solved by Riemann sums. For example, in the case of an exponential distribution, Lemma~\ref{lem:branchexp} shows an analytically tractable solution where the exponential distribution has rate $\theta$. In this case,
\begin{equation}
\label{eq:entropic_dist}
     \mathbb{E}(D^S_{ij}(d_{ij})) =2\bigg(\int_0^{d_{ij}/2}\bigg[e^{-\theta t}(S'(t) + \theta S(t)) + \int_0^{t}S(s)\theta^2 e^{-\theta s}ds\bigg] dt\bigg)
\end{equation}
and hence, the entropic likelihood is
\begin{equation}
    \ell_S(\mathcal{T}) = -2N\sum_{i\neq j}\bigg\{2^{-e_{ij}}\bigg(\int_0^{d_{ij}/2}\bigg[e^{-\theta t}(S'(t) + \theta S(t)) + \int_0^{t}S(s)\theta^2 e^{-\theta s}ds\bigg] dt\bigg)\bigg\}
\end{equation}
This is the likelihood that is used throughout the rest of the paper.

Before moving on to justify the use of this likelihood, it is informative to make two brief notes about its behaviour.
\subsubsection{Comparison to Felsenstein's likelihood}
The approximation (Equation \ref{eq:bmeobj}) means we use the data differently from Felsenstein's likelihood. Rather than using the data to directly calculate likelihoods via $p(g,\mathcal{U})$, the data are now used to create an inter-taxa distance matrix, from which the times $t_i$ are calculated.

This different application of the data means that the full dataset only needs to be used once in the optimisation process, as the distance matrix will remain the same throughout, independently of the topology under consideration. When the number of sites, $N$ is much larger than the number of taxa $n$ (which is often the case, particularly in phylogenomics) this results in a substantial saving in computational cost, as after a single pre-processing step, the dataset effectively reduces from size $Nn$ to size $n^2$.
\subsubsection{Connection to the classical balanced minimum evolution objective}
Defining $K = \sum_{a \neq b}\pi_a Q_{ab}$ to be the expected instantaneous substitution rate, assuming that
\begin{equation}
    \frac{K}{\theta} << 1
\end{equation}
and defining $\kappa := \min_{i,j}\bigg\{\theta d_{ij}\bigg\}$, one can show that the approximation
\begin{equation}
\label{eq:bmeapprox1}
    \frac{1}{2}\mathbb{E}(D^S_{ij}(d_{ij})) \approx \frac{1}{2}\mathbb{E}\bigg\{D_{ij}^S\bigg(\frac{\kappa}{\theta}\bigg)\bigg\} + \bigg(\frac{\theta d_{ij}}{2} - \kappa\bigg)\int_0^{\infty}\theta S(s)e^{-\theta s}ds
\end{equation}
has a percentage error of approximate leading order $\mathcal{O}(\log(\frac{K}{\theta})^{-1})$. This is formalised in Theorem~\ref{th:bme} in the Appendix. 

The assumption that $\frac{K}{\theta} << 1$ may seem contrived. However, noting that branch lengths are of length $\mathcal{O}(\frac{1}{\theta})$, the expected number of substitutions per site on a branch is $\mathcal{O}(\frac{K}{\theta})$. If this number is not small, then the sequences at each leaf node will be largely uncorrelated, regardless of the distance between them in the tree. Thus, this result shows BME is a good approximation to our model provided that there is a small number of mutations per branch. Note that $\frac{K}{\theta}$ is a non-dimensional quantity and independent of any reparametrisation of time.

\begin{figure}[htbp]
    \centering
    \includegraphics{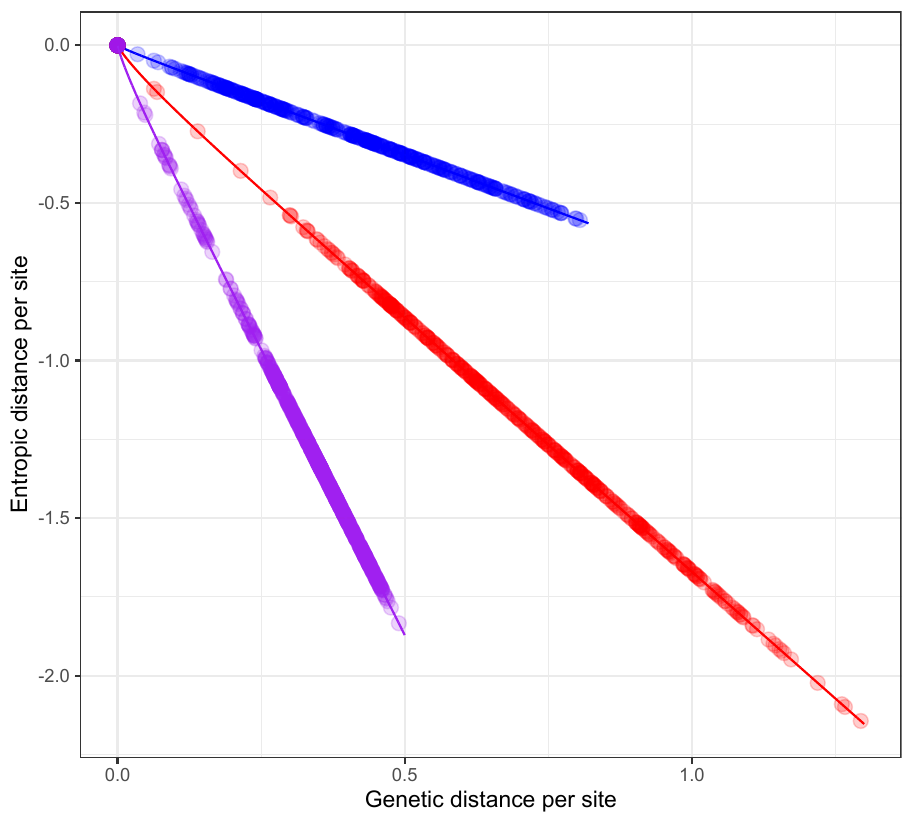}
    \caption{\textsb{Genetic distances $d_{ij}$ against corresponding entropic distances $\mathbb{E}(D^S_{ij}(d_{ij}))$ for three empirical datasets (see Table~\ref{tab:data}, DS1 is blue, DS2 is red and DS3 is purple).} For all datasets we see a strong linear relationship and correlation nearly equal to 1. Note that non-linearity tends to exist close to zero, where there is less data.}
    \label{fig:entropy}
\end{figure}

Using this result, we can note that (\ref{eq:bmeapprox1}) can be written as
\begin{equation}
\label{eq:approxbme}
    \mathbb{E}(D^S_{ij}(d_{ij})) \approx a + bd_{ij}
\end{equation}
for some constants $a$ and $b$. Then, the entropic likelihood (\ref{eq:bmeobj}) becomes (using the Kraft Equality~\citep{Parker1999-aa})
\begin{align}
    \ell_S(\mathcal{T}) &= -b\sum_{i \neq j}2^{-e_{ij}}d_{ij} + a\sum_{i \neq j}2^{-e_{ij}}\\
    &=  -b\sum_{i \neq j}2^{-e_{ij}}d_{ij} + a
\end{align}
which is a linear function of the classical BME objective and hence, in particular, the optimal tree under this objective will be the optimal BME tree. A visualisation of how linear this relationship is for three datasets (see Table~\ref{tab:data}) is shown in Figure~\ref{fig:entropy}.

\begin{table*}[h!]
\centering
\small
\caption{Evaluation datasets}
\begin{tabular}{llrrll}
\toprule
\textsb{Dataset} & Reference & \textsb{\# Sites} & \textsb{\# Taxa} & \textsb{Type} & \textsb{Taxonomic rank} \\
\midrule
DS1  &~\citep{hedges1990} & 1,949 & 27 & rRNA (18S) & Tetrapods \\ 
DS2  &~\citep{garey1996} & 2,520 & 29 & rRNA (18S) & Acanthocephalans \\ 
DS3  &~\citep{yang2003} & 1,812 & 36 & mtDNA & Mammals; mainly Lemurs\\ 
\bottomrule
\end{tabular}
\label{tab:data}
\end{table*}

Note that the value of $b$ can be written as 
\begin{equation}
   b =  \bigg(d_{ij}\theta\bigg)\bigg(\int_0^{\infty}\theta S(s)e^{-\theta s}ds\bigg) =\bigg(\mathbb{E}(\text{branches between $i$ and $j$})\bigg)\bigg(\mathbb{E}(\text{entropy on a branch})\bigg) 
\end{equation}
and hence, the new distances form a simple approximation of the entropy between each pair of taxa.

\subsection{Justification: Analytical and Empirical comparison to Felsenstein's likelihood}
As we show in this section, our entropic likelihood is similar, though not exactly equivalent, to the classical Felsenstein's likelihood. In particular, we demonstrate that the two are highly linearly correlated, meaning that, using an empirically calculated scaling factor, our entropic likelihood can be used to approximate sampling under Felsenstein's likelihood. We do this by deriving our likelihood directly from the expected Felsenstein's likelihood, and analysing the impact of each required approximation. 
\subsubsection{Setup}
Consider the expected Felsenstein's likelihood of a single-site tree $\mathcal{U}$ with branch lengths $\boldsymbol{b}^*$ given by balanced minimum evolution. This is equal to the entropy of the leaf nodes when the mutation process is realised on the true tree $\mathcal{T}$ with the true branch lengths $\boldsymbol{\beta}$ and then the resulting leaf sites are mapped to their equivalents in $\mathcal{U}$. 

Using $X$ to denote the set of leaf nodes and, later, $Y$ to denote the set of internal nodes, we define $H(X|\mathcal{T},\boldsymbol{\beta})$ to be the Felsenstein's likelihood.
\subsubsection{Using all nodes}
To reduce notation, we omit the conditional dependence of $H$ on $\boldsymbol{\beta}$ and $\mathcal{T}$ in this section.

As we vary $\mathcal{U}$ (and therefore also $\boldsymbol{b}^*$), we expect Felsenstein's likelihood $H(X)$ and the likelihood on all nodes, $H(X,Y)$ to be closely related to each other. Noting that
\begin{equation}
    H(X,Y) = H(X) + H(Y|X)
\end{equation}
the difference between them will be determined by the amount of information that the leaf nodes provide about the internal nodes.

In trees with a very low number of expected mutations per site (over the whole length of the tree), we expect that $H(X,Y) \approx H(X)$, as one can (with high probability) infer the states at the internal nodes from the leaf nodes - essentially using the fact that sites are extremely unlikely to mutate twice or more.

Conversely, in trees with a very high number of expected mutations per site, the sites at each node are approximately independent of each other (and therefore simply independent realisations of the stationary distribution of a continuous time Markov Chain mutation chain). In this case, as there are approximately the same number of leaf and internal nodes, we expect $\frac{1}{2}H(X,Y) \approx H(X)$.

In between these two extremes where there are intermediate numbers of expected mutations per site (where most real data exist), when we consider trees of similar topologies and branch lengths (and therefore $H(Y|X)$ varies slowly) it seems reasonable to approximate
\begin{equation}
\label{eq:approx1}
    H(X) \approx \kappa_1 H(X,Y)
\end{equation}

for some fixed $\kappa_1 \in [\frac{1}{2},1]$. That is, we expect the fully-observed likelihood to be approximately proportional to Felsenstein's likelihood.
\subsubsection{Varying the sampling distribution}
To recover our original new likelihood $\ell(\mathcal{U})$ from (\ref{eq:entexact}), we must make the approximation
\begin{equation}
\label{eq:approx2}
   H(X,Y|\mathcal{T},\boldsymbol{\beta}) \approx  \kappa_2H(X,Y|\mathcal{U},\boldsymbol{b}^*) = \kappa_2L(\mathcal{U})
\end{equation}
The effect of this approximation is to change the assumed distribution of the sites when calculating the entropy. $ H(X,Y|\mathcal{T},\boldsymbol{\beta})$ assumes the true distribution but, in general, this is unknown. When considering the tree $(\mathcal{U},\boldsymbol{b}^*)$, we hence instead perform our calculation \textit{as if this tree were the correct tree}.

In Lemma~\ref{lem:k2}, we justify this approximation and show that we expect $\kappa_2 \in [1,2]$. In essence, the primary reason for this is that as we consider trees $\mathcal{U}$ which are further from the true tree $\mathcal{T}$, their entropy will be higher and therefore there is a double effect on $H(X,Y|\mathcal{T},\boldsymbol{\beta})$ - it increases both because $\mathcal{U}$ is a higher entropy tree, and because the true sequence distribution is diverging from that given by $\mathcal{U}$.

Unlike the previous approximation (\ref{eq:approx1}), where we essentially just change the likelihood scale, the approximation (\ref{eq:approx2}) does add meaningful error and means that the correlation between the entropic and Felsenstein's likelihood is not as strong as, for example, the correlation between the entropic likelihood and the BME objective - which is virtually the same. However, we have found that these errors are not prohibitive, as we show in the subsequent section, where we make simple empirical comparisons between the entropic and Felsenstein's likelihood.

\subsubsection{The entropic likelihood}
The final approximation comes from the use of our entropic distance matrix to approximate the value of $L(\mathcal{U})$. The validity of this approximation depends on how well our heuristic tree branching process model describes the true process of tree creation, and can change the gradient of the linear relationship between the entropic likelihood and Felsenstein's likelihood.

To examine the effect of this, suppose that we have two branching models, A and B which determine the construction of a tree. In model A, we suppose that, for a uniformly chosen branch in the tree of length $T_i$, we have $\mathbb{E}(S(T_i)) = s_A$ and $\mathbb{E}(T_i) = \tau_A$, with similar definitions for model $B$.

Then, for trees which are a sufficiently large distance $\Delta$ apart, we expect entropic distances $S_{\cdot}$ to be approximately equal to the average entropy on a branch, $s_{\cdot}$, multiplied by the number of branches, approximately $\frac{\Delta}{\tau_{\cdot}}$. That is,
\begin{equation}
    S_A = \frac{s_A}{\tau_A} \Delta \quad \text{and} \quad S_B = \frac{s_B}{\tau_B} \Delta
    \label{eq:distribution_difference}
\end{equation}
Thus, $S_A$ and $S_B$ are linearly related, but the ratio between them (and therefore the ratio between the entropic likelihoods given by the two tree construction models) will not necessarily be 1. While it is readily possible to re-estimate model parameters for each tree, this will not resolve all the previous sources of error in our approximation. We therefore propose calibrating our entropic likelihood against. Felsenstein's likelihood via a linear scaling with gradient $m$.

\subsubsection{Analytical summary}
Thus, we see that our entropic likelihood and Felsenstein's likelihood should be approximately linearly related. This is an important result, not only for justifying the validity of the entropic likelihood, but also in providing an explanation for the utility of BME - we can use the approximate linear relationship between BME and the entropic likelihood to show that we also expect BME to be approximately linearly related to Felsenstein's likelihood. This theoretical finding opens the door for approximate Bayesian inference using a distance matrix and BME, simply by performing a calibration to estimate the linear scaling coefficient $m$.  It is critical to note that, even after scaling, the entropic likelihood will not exactly match Felsenstein's likelihood. However, given linearity, the two likelihoods will be of similar magnitudes for a given tree.

\subsubsection{Illustrative empirical comparisons to Felsenstein's likelihood}

First, we illustrate through a simple empirical example that a linear approximation is reasonable. As highlighted in the motivation, we do not expect this likelihood to perform as well as Felsenstein's likelihood because we ignore information provided about the topology $p(\mathcal{D}|\mathcal{U})$. In contrast, Felsenstein's likelihood incorporates this information through a post-order tree traversal and marginalisation. However, using our entropic likelihood is a reasonable choice when computational tractability prohibits the use of Felsenstein's likelihood.

\begin{figure}[htbp]
    \centering
    \subfloat[]{\includegraphics[trim={0 2cm 25cm 1.5cm}, clip, width=0.48 \linewidth]{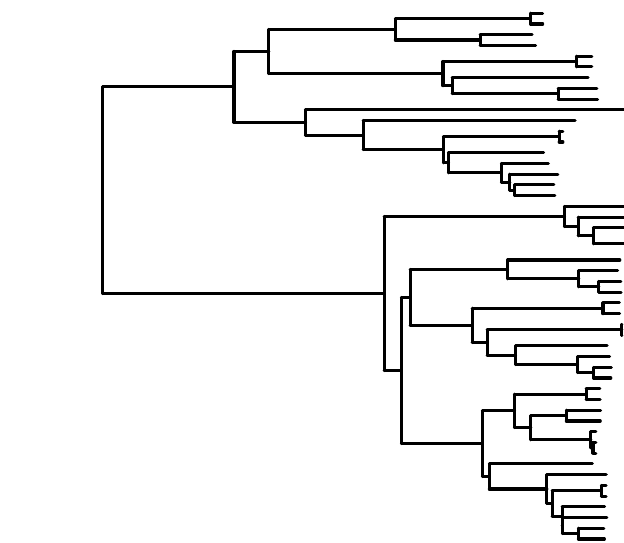}\label{fig:fig1a}}
    \\
    \subfloat[]{
        \includegraphics[trim={11.5cm 0 12.5cm 1cm}, clip, width=0.48\linewidth]{Entropic_likelihood.pdf}}
    \subfloat[]{
        \includegraphics[trim={23.5cm 0 0.5cm 1cm}, clip, width=0.49\linewidth]{Entropic_likelihood.pdf}}
    \caption{Comparison of entropic and Felsenstein's likelihoods assuming exponential branch lengths. \textsb{(a)} Example single true tree from simulation. \textsb{(b)} A comparison of the likelihoods of suboptimal trees inferred from data simulated through the true tree. Black dots are suboptimal trees generated by performing random SPR moves from the best estimated tree, and red dots are entirely random trees. \textsb{(c)} A comparison of the likelihoods for the best tree across 2000 simulated alignments. }
    \label{fig:entropic_vs_fels}
\end{figure}

To show the limitations of our entropic likelihood, we simulate a biologically realistic alignment from a known tree. Our tree simulation follows a birth death process with 50 species and with $\lambda=0.5$, $\mu=0.1$, $\rho=1$ and time since origin (TMRCA) of $65$ (reflecting many major radiations since the K-Pg transition event). On this birth-death process, we simulate white noise with a mean rate and rate variation of 0.05 according to an exponentiated white noise process, resulting in samples that are distributed log normally. A birth-death tree with noise is considered the true tree $\mathcal{T}$ and is created using the \texttt{TreeSim}~\citep{Stadler2011-dx} and \texttt{NELSI}~\citep{Ho2015-up} packages in R. To simulate an alignment down this tree we use the \textit{seqSim} function in \texttt{phangorn}~\citep{Schliep2011-zu} in R assuming 4 nucleotides and no gaps or unknown bases. We sample the six possible transition rates as well as the for possible base frequencies from a Dirichlet distribution with $\alpha_i=5$ $ \forall i$. Finally, we assume a Gamma shape of $1$ with $4$ categories and sample rates using the \textit{discrete.gamma} function in \texttt{phangorn}. Finally, we assume a sequence length of $5,000$ base pairs. An example of a resultant tree is shown in Figure~\ref{fig:entropic_vs_fels}. For all inferences, we assume a Jukes-Cantor substitution model and therefore are performing inference under model misspecification. We compare our likelihood to Felsenstein's likelihood for alignments simulated as described above. Our likelihood is:
\begin{equation}
    \ell_S(\mathcal{T}) = -N \sum_{i,j} \mathbb{E}(D^S_{ij}(d_{ij})) 2^{-\epsilon_{ij}}
\end{equation}
Assuming exponential branch lengths ($f(t)\sim\text{Exponential}(\theta)$) with the distance matrix found by solving the renewal equation
\begin{align}
  \frac{1}{2}\mathbb{E}(D^S_{ij}(d_{ij})) := h(\tau)  = (1-F(\tau))S(\tau) + \int_0^{\tau}(h(\tau - t) + S(t))f(t)dt
\end{align}
The rate of the exponential distribution was specified as $\hat{\theta}  = \frac{n-2}{L}$ where $L$ is the tree length (this formula is derived in Lemma \ref{lem:theta}). Using Felsenstein's likelihood to estimate an optimal tree was performed using PAML~\citep{Yang2007-ez}, implemented in the \texttt{phangorn} library~\citep{Schliep2011-zu} in R.

Simulating $2000$ random alignments and optimising our likelihood and Felsenstein's likelihood and examining one minus the normalised Robinson-Foulds distance~\citep{robinson1981} to the known true tree results in a median topological accuracy of $0.9787$ $[0.9361-1]$ for both our entropic likelihood approach and Felsenstein's likelihood approach. Examining the likelihoods for these $2000$ random alignments, we see a strong linear relationship between the optimal likelihoods (see Fig.~\ref{fig:entropic_vs_fels}, right) with $m=0.92$. Therefore, for the optimal tree, the scaling is close to one. Picking one single fixed tree and alignment, we can explore the relationship between our entropic likelihood and Felsenstein's likelihood for suboptimal trees given the alignment. Generating $2000$ trees by subtree-prune-and-regraft (SPR) operations on the optimal tree, where the number of operations is randomly drawn uniformly from integers between $1$ and $50$ i.e. maximum 50 sequential SPR moves also results in a strong linear relationship ( Fig.~\ref{fig:entropic_vs_fels}, middle black points). Sampling $2000$ entirely random trees and evaluating both likelihoods (Fig.~\ref{fig:entropic_vs_fels}, middle) further interpolates the linear trend. We note that when examining suboptimal trees $m=0.508$, and is therefore not a close scaling. However, the linear relationship that our theory suggests holds for trees close to the optimal tree, all the way to entirely random trees. 

\begin{figure}[htbp]
    \centering
    \includegraphics[scale=0.45]{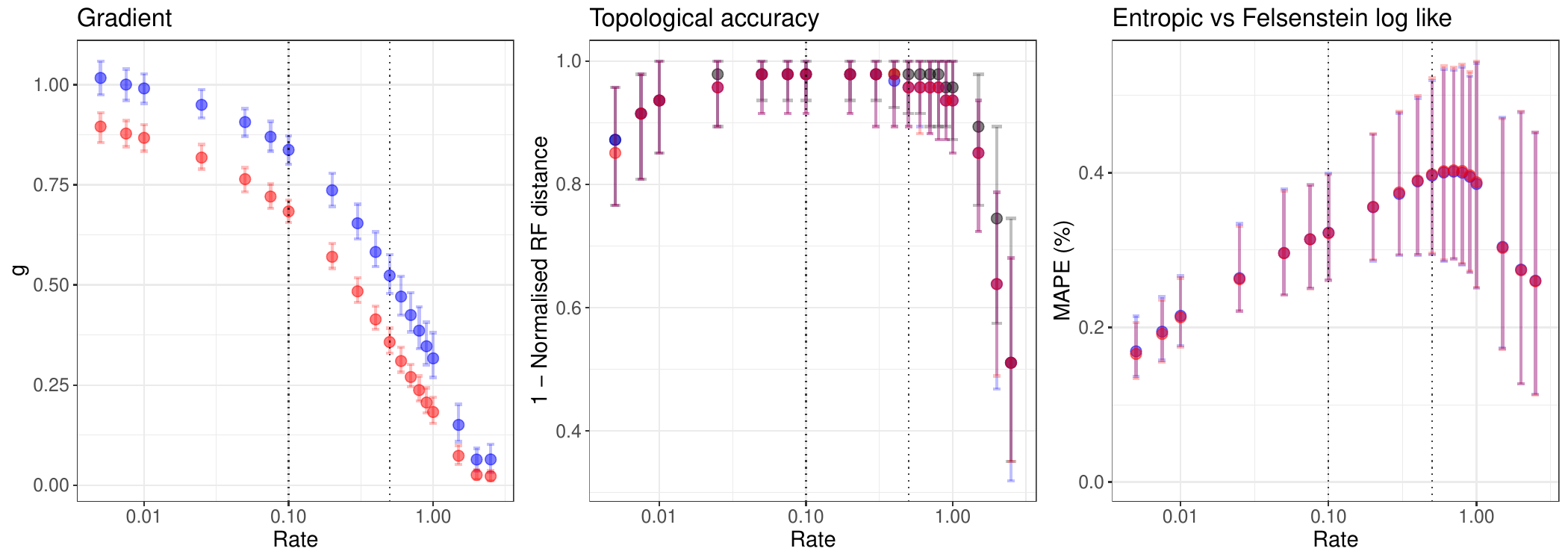}
    \caption{\textsb{Performance of the entropic likelihood across evolutionary rates.} Left: The variation of the scaling between the entropic and Felsenstein's likelihood as a function of rate, where red indicates the entropic distance estimated using Equation~\ref{eq:entropic_dist} assuming exponentially distributed branch lengths. Blue points use the same procedure, but the branching rate $\theta$ is re-estimated for each tree and a new entropic distance found. Middle: Comparison of topological accuracy for Felsenstein's likelihood in black, entropic distance in red, and entropic distances when restimating $\theta$ in blue. Right: The mean absolute percentage error between Felsenstein's likelihood and a linear model of entropic distance. The percentage error is on a log likelihood scale. Black dotted lines show where most empirical data exist~\citep{Klopfstein2017-uv}.}
    \label{fig:error}
\end{figure}

Next we show how the gradient, $m$, between the Felsenstein and entropic likelihoods varies between 0 and 1 across alignments with different rates. As we have noted from our discussion of the $\kappa_1$ scaling parameter in~\eqref{eq:approx1}, our theory suggests with low rates we expect a gradient, $m$,  of around one when compared to Felsenstein's likelihood, and with high rates we expect $m$ to decrease. This decrease is also caused by the entropic likelihood approximation, as higher rates lead to entropy being less closely linear. Simulating trees using the same procedure as above with 50 taxa and 5,000 sites, we explored how rate variation affects our approximation via $m$. Simulated trees were rescaled following the procedure outlined in~\citep{Klopfstein2017-uv}. Briefly, varying evolutionary rates were set as the root-to-tip divergences. We rescaled root-to-tip distances to vary between a very wide range of $\{0.005,2.5\}$~\citep{Klopfstein2017-uv}, and for each rate simulated 500 trees to estimate uncertainty and for each tree we explore the distribution around the optimal tree via SPR changes close to the optimal tree, where the number of sequential random SPR changes is drawn from $\sim \text{Poisson}(1)+1$. To mitigate the effects of a misspecified tree construction model, we estimate $\theta$ for each new SPR tree which, for the exponential case, is available in closed form as $\hat{\theta}  = \frac{n-2}{L}$ where $L$ is the length of the tree found as the BME objective. This helps reduce the error for trees far from the true tree (which through classical BME theory, will have larger values of $L$), but does not reduce the impact of the overall model misspecification (an impact, which as discussed, will increasingly affect the gradient as the substitution rate grows). As shown in Figure~\ref{fig:error} (left), re-estimating $\theta$ does result in a coefficient $m$ that is close to 1 when rates of substitution are low. However, as rates increase and multiple substitutions happen and multiple sites, the estimates of $m$ become progressively worse. The topological accuracy of the best tree however remains excellent and comparable to Felsenstein's likelihood for all rates (Fig.~\ref{fig:error}, middle). Similarly, the mean absolute percentage error between Felsenstein's likelihood and a linear model with the entropic likelihood does grow with rate, but not substantially, and considering the log scale, the linear approximation is good with up to half a percentage point of error.

In summary, the linearity of our entropic likelihood with Felsenstein's likelihood is a useful property, and justifies the previous use of BME in likelihood proposals e.g.~\citep{Guindon2010-ip,stamatakis2014}. However, because there is no guarantee that the gradient is 1, and indeed we expect it not to be for real data~\citep{Klopfstein2017-uv},  Bayesian distance-based MCMC inference cannot be performed without a linear calibration. A solution we utilise is to simply perform a calibration of the entropic likelihood against Felsenstein's likelihood to estimate the gradient correction. This calibration only requires a few hundred trees and can be performed as an average across subsets of taxa for large numbers of sequences, or as a subset of the number of sites for a very large number of sites (e.g. a whole genome). The resultant posterior is of course with respect to our approximate likelihood and will not converge to Felsenstein's likelihood, but a judicious choice of data and model to ensure a good distance approximation will make it close. We note that our theory and these illustrative simulations show that the BME objective can be treated as a likelihood that needs to be linearly calibrated against a choice of phylogenetic likelihood. For example, for the example datasets in Table~\ref{tab:data}, $m$ using entropic likelihoods are estimated as $m=\{0.9119,0.7882,0.7635\}$. One can also calculate the gradients between standard BME and Felsenstein's likelihood, which are $\{8264,4731,4730\}$. 

\subsubsection{A Generalized Bayesian perspective}

Our theoretical analysis above shows that we expect that for most real-world data, our entropic likelihood, $\ell$, will be linearly related to the true Felsenstein's log-likelihood $\ell_F$ 
\begin{equation}
    \ell_F(\mathcal{U}) = c + m\ell_u(\mathcal{U}) + \epsilon(\mathcal{U})
\end{equation} 
for some parameters $m$ and $c$. Here, for clarity, we use a subscript $u$ to show that $\ell_u$ is our unscaled entropic likelihood. As discussed, this relationship arises primarily from the lack of marginalisation up a given tree and only considering leaf to leaf paths. However, once we suitably scale $\ell$ to be $\ell_r$, we expect
\begin{equation}
    \ell_F(\mathcal{U}) = \ell_r(\mathcal{U}) + \epsilon(\mathcal{U})
\end{equation} 
and there is reason to believe that model misspecification of this rescaled entropic likelihood is not of substantial concern.  Certainly, in our examples (e.g. Fig.~\ref{fig:entropic_vs_fels}), we empirically observe that the error term $e^{-\epsilon}$ is much smaller than the average distance between trees, particularly when we restrict our attention to the trees topologically close to the optima. 

To explore this more precisely, we follow the method in generalised Bayesian analysis \citep{Bissiri2016-nl} and consider assigning our likelihood a weight $w$, with the remaining weight $(1-w)$ being given to a null model - that is, a single constant log-likelihood $\ell_C$ across all trees. We therefore consider a likelihood of
\begin{equation}
    wL(\mathcal{U}) + (1-w)L_C
\end{equation}
As shown in \citep{Bissiri2016-nl}, if the scaled entropic \emph{likelihood} $L(\mathcal{U})$ is misspecified no meaningful prior can be set, and it is preferable to consider $L(\mathcal{U})$ as a loss function and the parameter $w$ ensures Bayesian updates remain coherent\citep{Bissiri2016-nl}.

To illustrate that our rescaled entropic likelihood results in $w \approx 1$ we simulate 1,000 unique, 50 taxa trees and alignments (5000 sites each) from a birth death process as previously outlined using a rate of 0.2\citep{Klopfstein2017-uv}. We then sampled 1,000 unrooted trees as random SPR moves from the optimal tree, with the number of SPR changes randomly distributed as $\sim \text{Poisson}(1)+1$. Using this random sample of trees, we can calculate the rescaling to get $\ell_r$ via linear regression. 

Given our sample, we can calculate approximately normalised tree likelihoods under our different models, which we denote by 
\begin{equation}
    p_{*}(\mathcal{U})=\frac{L_{*}{U}}{\sum_\mathcal{U} L_{*}(\mathcal{U})}
\end{equation}
Note that because of the comparative closeness of the sampled trees to the optimum, and therefore their relatively high likelihoods, we use this definition of $p_C$ to give a fair comparison of our likelihood's optimal weight. Setting $p_C$ to be simply 1/(number of trees) gives a far less accurate null model and therefore better performance when our model is more highly weighted.

We can then estimate the Kullback–Leibler divergence between the distribution from Felsenstein's likelihood and our unscaled entropic likelihood as
\begin{align}
     \mu = \sum_{\mathcal{U}} p_f(\mathcal{U}) \text{log}\left(\frac{p_f(\mathcal{U})}{w p_*(\mathcal{U}) + (1-w)p_c(\mathcal{U})}\right)
\end{align}
Where $p_*(\mathcal{U})$ is evaluated for both our standard and scaled entropic likelihood.
\begin{figure}
    \centering
    \includegraphics[width=1\linewidth]{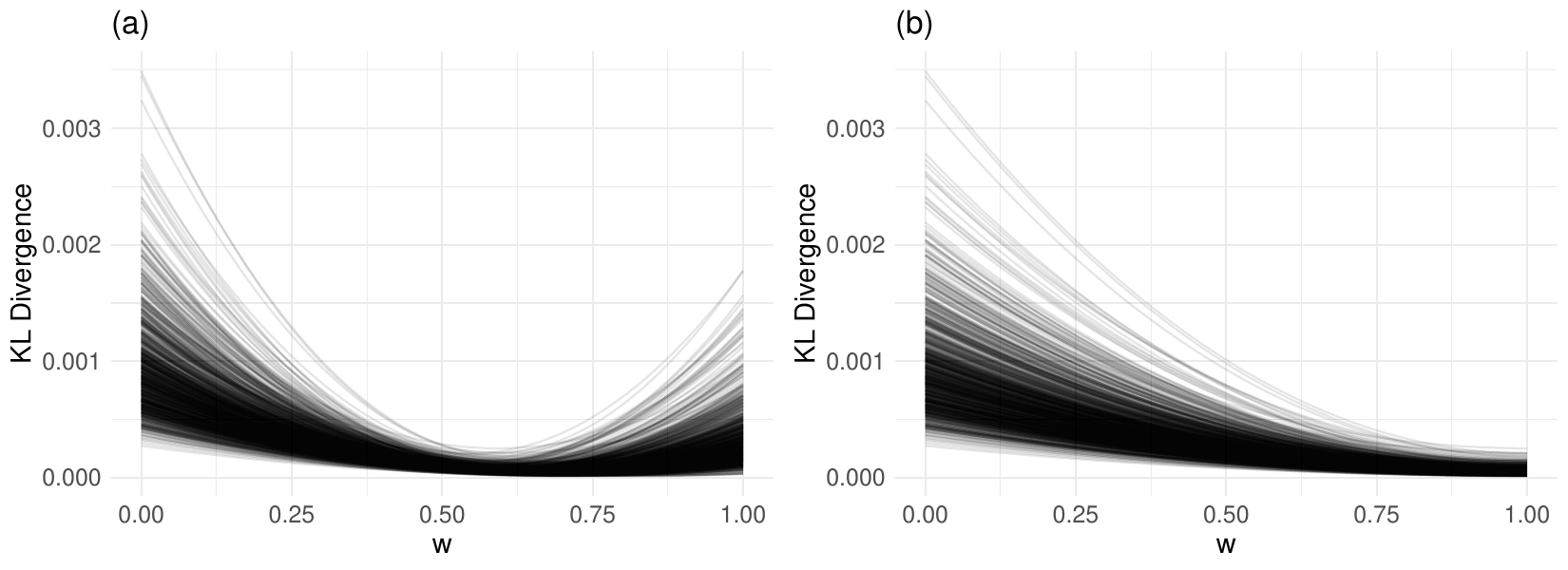}
    \caption{Kullback-Leibler divergence across weight values for our likelihood, for 1,000 50 taxa alignments with 5,000 sites, generated from a birth-death process. The Kullback-Leibler divergence is generated from a random sample of trees perturbed from the optimal tree via SPR trees. (a) shows the standard entropic likelihood which suffers from misspecification due to a variety of approximation errors, (b) shows the calibrated entropic likelihood which does not suffer from substantial misspecification}
    \label{fig:KLdivergence}
\end{figure}
In Figure~\ref{fig:KLdivergence} we examine the KL divergence and find that our entropic likelihood does not have minimal KL divergence at $w=1$ suggesting model misspecification, and reinforces our earlier theory on the various errors in our approximation. However, our rescaled calibrated entropic likelihood does achieve a minimal KL divergence at $w=1$ for all simulations and therefore we can conclude that this simulated example does not suffer from substantial misspecification and that our simple scaling is sufficient to correctly specify the model. These simulations are not definitive, but a theoretical asymptotic analysis is not possible given the complexities of both likelihoods. This analysis simply provides empirical reassurance to the theory we have already outlined. 

\subsection{Implementation procedures}

To estimate the entropic distance matrix, we first require a sequence alignment, from which we estimate the standard distance matrix $D$. This can be easily achieved for common reversible Markov substitution models. Next, we need to specify a branch length distribution $p(\beta)$. This can be chosen \textit{a priori} or taken from a reasonable estimate from a tree. In our examples, we first find an (approximately) optimal BME tree, calculate the least squares solution to the branch lengths, and then use these lengths to inform our distribution. Assuming an exponential distribution, $p(\beta) \sim \text{Exp}(\lambda)$ where $\lambda$ is the reciprocal of the mean across all branch lengths in the tree (the maximum likelihood solution). Other distributions can readily be used. Given a parametric choice for $p(\beta)$, the density and cumulative distribution functions are defined and the renewal equation (\ref{eq:renewal_eq}) can be solved. Solving this equation for values of $D$ results in a new entropic distance matrix $D^s$. Given $D^s$, we can choose a sampling of $N$ trees $\{\mathcal{U}_1,\ldots, \mathcal{U}_N\}$ and regress our likelihood against another phylogenetic likelihood such as Felsenstein's likelihood to obtain an estimate of the scaling $m$. We note, that due to the theory we introduce showing the linearity between the entropic distance and the genetic distance, this calibration can be done directly on the standard BME objective (taken from $D$) rather than the entropic likelihood (taken from $D^s$) with only small additional error. However, the resultant likelihood will be orders of magnitude different when performing the regression from standard BME, which can introduce numerical sensitivity.

\subsubsection{Run time comparison}
 To illustrate the improvement in utilising our entropic likelihood instead of Felsensteins, we perform an illustrative run time comparison analysis. We simulate trees using a birth-death process as previously described, and combinatorially vary both the number of taxa and the number of sites. Felsenstein's likelihood was calculated using RAxML-NG~\citep{kozlov2019} where a fixed topology and a Jukes-Cantor model were used, but branch lengths were estimated. Our entropic likelihood was calculated using the \texttt{APE}~\citep{paradis2019} function \texttt{cophenetic.phylo}. The computation of the distance matrix was not included in the run times as this only needs to be calculated once before inference. We observe that Felsenstein's likelihood scales orders of magnitude slower than our entropic likelihood, and this is particularly evident as the number of sites increases. For our genomic example on birds below where the length is 60 million sites, evaluation of Felsenstein's likelihood for a given tree would be prohibitively slow.

\setlength{\tabcolsep}{2.5pt}
\begin{table}[htbp]
\centering
\small
\begin{tabular}{l*6r@{\hspace{0.5cm}}*6r}
\toprule
                        & \multicolumn{6}{c}{\textsb{Balanced Minimum Evolution}} & \multicolumn{6}{c}{\textsb{Felsenstein's Likelihood}} \\ 
                        & \multicolumn{6}{c}{\textsb{Number of Sites $\rightarrow$}} & \multicolumn{6}{c}{\textsb{Number of Sites $\rightarrow$}} \\ 
\textsb{Number of taxa $\downarrow$} & \textsb{100} & \textsb{500} & \textsb{1k} & \textsb{10k} & \textsb{50k} & \textsb{100k} & \textsb{100} & \textsb{500} & \textsb{1k} & \textsb{10k} & \textsb{50k} & \textsb{100k} \\
\midrule
\textsb{50}             & 0.002        & 0.006        & 0.001          & 0.000               & 0.000               & 0.001           & 0.040         & 0.040         & 0.040           & 0.150            & 0.630            & 0.990            \\
\textsb{100}            & 0.001        & 0.001        & 0.001          & 0.001           & 0.001           & 0.001           & 0.040         & 0.050         & 0.060           & 0.320            & 1.410            & 3.010            \\
\textsb{200}            & 0.002        & 0.002        & 0.003          & 0.003           & 0.003           & 0.003           & 0.040         & 0.060         & 0.100           & 0.580           & 2.730            & 5.350           \\ 
\textsb{500}            & 0.009        & 0.009        & 0.008          & 0.018           & 0.019           & 0.018           & 0.060         & 0.120         & 0.180           & 1.400            & 6.640            & 14.990           \\ 
\textsb{750}            & 0.020         & 0.020         & 0.032          & 0.040            & 0.019           & 0.038           & 0.070         & 0.150         & 0.260           & 2.040            & 11.670           & 23.310           \\ 
\textsb{1,000}          & 0.039        & 0.045        & 0.046          & 0.051           & 0.077           & 0.079           & 0.080        & 0.180         & 0.330           & 2.730            & 15.520           & 31.340         \\ 
\textsb{2,500}          & 0.266        & 0.313        & 0.255          & 0.263           & 0.427           & 0.395           & 0.170         & 0.405         & 0.770           & 6.840            & 42.330           & 83.860           \\ 
\bottomrule
\end{tabular}
\caption{Comparison of run times (seconds) for balanced minimum evolution and Felstensteins likelihood for different numbers of taxa and sites.}
\label{tab:combined}
\end{table}
\setlength{\tabcolsep}{6pt}

\bigskip
\section{Results and discussion}

\subsection{Bayesian distance-based inference on standard benchmark datasets}

\begin{figure}
    \centering
    \includegraphics[scale=0.8]{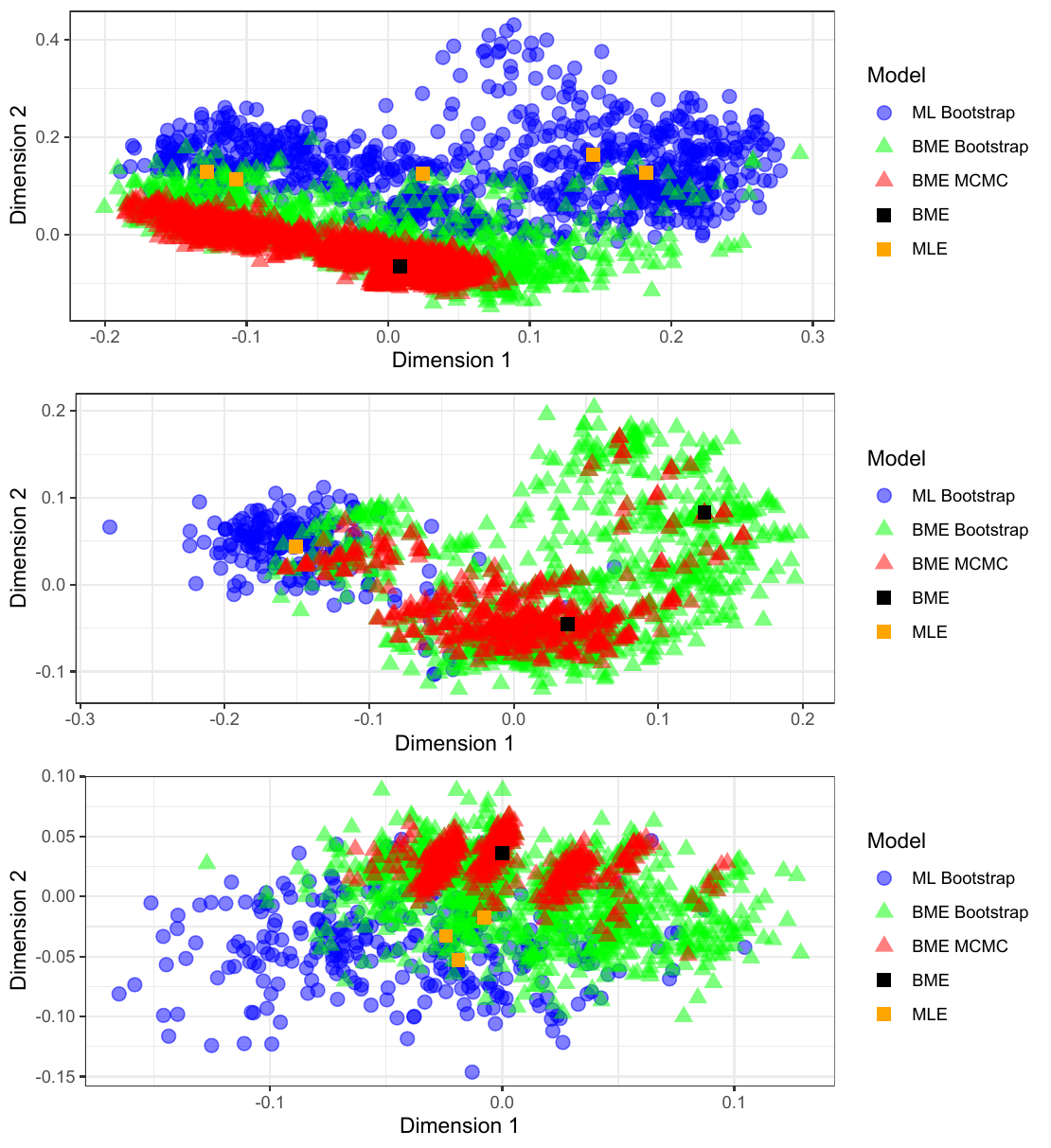}
    \caption{For the set of unique trees, the generalised Robinson-Foulds distance is calculated, and the distance matrix reduced by multidimensional scaling. A Jukes-Cantor model was used. ML Bootstrap: RAxML-NG~\citep{kozlov2019} bootstrap, BME bootstrap: FastME~\citep{lefort2015} bootstrap, BME MCMC is the method presented in this paper, and BME and MLE are the point estimates. To facilitate visualisation, only a random sample of 5000 trees is shown.}
    \label{fig:DS}
\end{figure}

To compare current phylogenetic inference approaches with the entropic likelihood, we started by using three classical benchmark datasets in phylogenetics. These included data on tetrapods~\citep{hedges1990} (27 taxa, 1949 sites), acanthocephalans~\citep{garey1996} (29 taxa, 2520 sites), and mammals~\citep{yang2003} (36 taxa, 1812 sites; hereafter DS1-3). For each of the three datasets, we performed phylogenetic inference using (i) maximum likelihood implemented in RAxML-NG~\citep{kozlov2019} optimisation with 100 different starts and 1000 bootstraps with the \texttt{autMRE} function (extended majority-rule consensus tree criteria with a cut-off of 0.03). From the 100 different starts, a subset of unique local minima was created. (ii) BME using FastME~\citep{lefort2015} with 1000 bootstraps. We also included (iii) the new continuous vector-based BME inference using GradME~\citep{penn2023leaping} to search for an optimal tree. We implement the entropic likelihood by running a Random Walk Metropolis-Hastings Markov Chain Monte Carlo for 20 million iterations, with 500,000 burn-in discarded and across 20 chains, thinning every 10 samples. Convergence was assessed by calculating the R-hat statistic on the chain log-likelihoods, as well as visual inspection of the like-likelihood trace plots. To guarantee linearity between likelihoods (see Methods), we calibrated the entropic likelihood to Felsenstein's likelihood using 1000 trees, perturbed from the best balanced minimum evolution tree with SPR$(x)$, where $x$ is the number of sequential SPR moves and $x\sim \text{Poisson}(1)+1$ - that is mostly small changes of one SPR from the best tree, but occasionally large changes. The results are visualised in Figure~\ref{fig:DS}. To maintain consistency with previous studies~\citep{Whidden2015-jn}, all analyses were performed on a Jukes-Cantor substitution model~\citep{jukes1969}.

To compare all optimal, bootstrap, and posterior trees across analyses, we calculated pairwise Robinson-Foulds distances across trees for the datasets enumerated in Table~\ref{tab:data}. To visualise these distances, we follow~\citep{Whidden2020-gp} and plot the first two components of a multidimensional scaling reduction of the distance matrix. For DS1 (see Fig.~\ref{fig:DS}a), a very large number of trees were sampled ($\sim$ 30,000) and we see the distance tree is relatively far from the maximum likelihood tree (around 0.5 normalised RF distance from the RAxML-NG modes). The bootstrap distributions from maximum likelihood and distance estimation overlap, as does the Markov chain Monte Carlo (MCMC) posterior. The MCMC entropic posterior is more concentrated but with considerable variation around the BME optimum. This narrower sampling of the posterior set of unique trees is expected~\citep{Holmes2003-cg}.

In DS2, the continuous tree search~\citep{penn2023leaping} found a better tree than the optimal FastME tree, which was the most similar to the RAxML-NG tree. In this dataset, only one RAxML-NG mode was found across all 100 runs, and the bootstrap (blue circles) is tighter than in DS1 (Fig.~\ref{fig:DS}b). The entropic MCMC posterior ranges across the BME bootstrap samples and includes trees very close to the best maximum likelihood tree. In DS2, FastME selected a tree which was suboptimal, while GradME found a tree with a smaller length that was closer to the maximum likelihood best tree. Reassuringly, the entropic MCMC samples around both of these modes, and a third mode that is close to the maximum likelihood tree. For DS3, we once again see that the entropic MCMC explores several modes that correspond closely to the BME bootstrap distribution, and samples trees close to one of the best maximum likelihood trees. Both the FastME and RAxML-NG bootstraps overlap considerably, and the distance between optimal trees is small (normalised RF distance $\sim 0.1$).

Overall, these analyses demonstrate that the entropic likelihood facilitates MCMC sampling, with the distribution of samples overlapping with a distance-based bootstrap. The number of unique topologies varies depending on how well-resolved the tree is based on a distance matrix; for DS1, over 30,000 unique topologies were found, for DS2, only 413 were found, while for DS3, only 382 were found. The correspondence between the distance-based posterior and the distance-based bootstrap validates the theoretical results above, suggesting that this likelihood can be viewed as BME with a suitable scaling. This means that by using the entropic distance matrix as opposed to a standard distance matrix (e.g., Jukes-Cantor distances), it is possible to perform Bayesian model-based inference using all the theories and methods already developed in BME. BME has a quadratic complexity and can scale to thousands of trees, opening the possibility of Bayesian inference on huge datasets with thousands of taxa and millions of sites. We note that we have only used a single distance matrix to estimate the posterior distributions, propagating uncertainty from the distance matrix will create a larger posterior distribution of unique trees.

\subsection{Bayesian inference on the Bird 10,000 Genomes (B10K) dataset}

To showcase the results that can be obtained from analyses using the entropic likelihood, we used data from the Bird 10,000 Genomes (B10K) project~\citep{feng2020,stiller2024}, an initiative that aims to generate representative draft genome sequences from all extant bird species. Here we analyse the release of 363 genomes representing 92\% of all bird families. We estimated pairwise genetic distance under a GTR+$\Gamma$ substitution model (general time-reversible model with gamma-distributed rate variation among sites) using the approach outlined in~\citep{yang2006}. Given free parameters for rates, $S\in\mathbb{R}_{\geq 0}^{6}$, frequencies, $\pi\in\mathbb{R}_{\geq 0}^{4}$ and $\sum \pi_i=1$, the time between two sequences $t_{ij}$, and genetic sequence data $\mathcal{G}$, the log-likelihood for the transitions between a pair of taxa $i$ and $j$ is
\begin{equation}
    \mathcal{L}_{ij}(\mathcal{G}|S,\pi,t_{ij}) = \sum_a \sum_b \kappa^{ij}_{ab} \log(P_{ab}(t_{ij}; S,\pi))
\end{equation}
where $\kappa^{ij}_{ab}$ is the number of $a \to b$ transitions from taxon $i$ to taxon $j$. We approximate the optimal parameters by minimizing the total negative log-likelihood (that is, the sum over $i$ and $j$ of $\mathcal{L}_{ij}$) using gradient descent in \texttt{Jax}~\citep{jax2018github}.

We considered two alignments, a smaller alignment, $A_1$, of 100,000 sites that was originally used to estimate a maximum likelihood tree in RAxML-NG~\citep{stamatakis2014,feng2020} and a much larger alignment, $A_2$, of 63.4 million sites from intergenic regions across the genomes as used for phylogenomics in this data set~\citep{stiller2024}, and for which maximum likelihood tree estimation is highly impractical. For both alignments $A_1$ and $A_2$, assuming a GTR+$\Gamma$ substitution model as above, distance matrices $D_1$ and $D_2$ could be estimated. Comparing these distance matrices to the RAxML-NG tree created from $A_1$, we find the distance matrix from the smaller alignment, $D_1$, yields a topological accuracy (one minus the normalised Robinson-Foulds distance) to the RAxML-NG tree of 91\% - that is, 91\% of all splits are shared between the trees. Using the distance matrix from the larger alignment $D_2$, the topological accuracy to the RAxML-NG tree increases substantially to 98\%. Critically, both distance matrices are highly linearly related (intercept 0.004911  and gradient 1.051129), and therefore we can perform a likelihood calibration, applying a calibration on $A_1$ to the more reliable distance matrix $D_2$. The calibration of the entropic likelihood to Felsenstein's likelihood was done on $A_1$ using 500 trees, perturbed from the best balanced minimum evolution tree with SPR$(x)$, where $x$ is the number of sequential SPR moves and $x\sim \text{Poisson}(1)+1$. 

\begin{figure}
    \centering
    \includegraphics[scale=0.18]{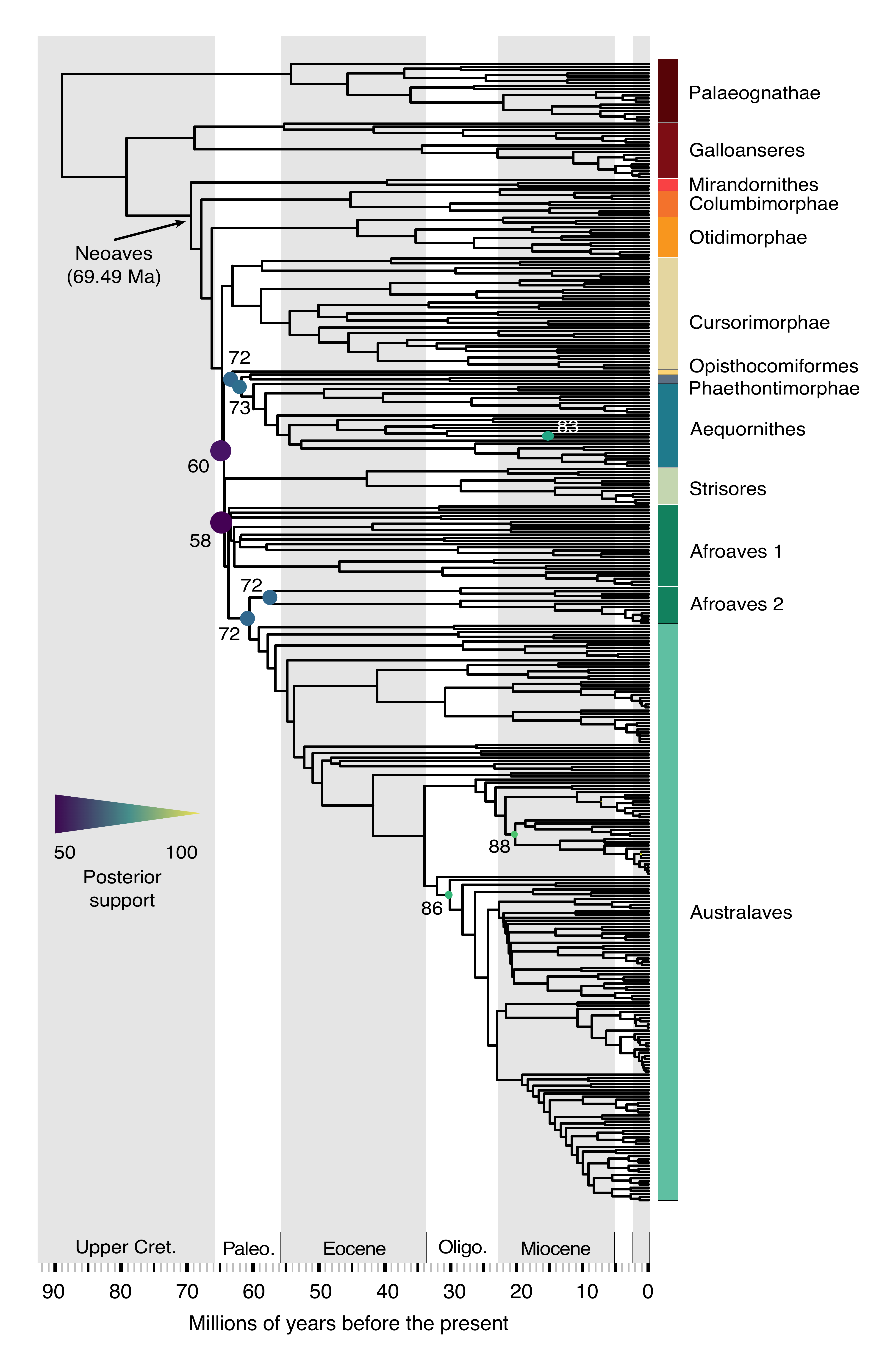}
    \caption{Maximum clade credibility tree with posterior node support for 363 bird taxa built from a distance matrix of $\sim 60$ million sites~\citep{feng2020,stiller2024}. The tree has a normalised Robinson-Foulds distance of $~0.02$ (8 splits)  from the maximum likelihood tree, and the posterior uncertainty in tree topology concentrates on the K-Pg boundary, in line with prior expectations~\citep{feng2020}. Molecular dating was performed for visual purposes on a subset of calibrations from the original study.}
    \label{fig:b10k}
\end{figure}

Our avian family-level Bayesian phylogenetic estimate is highly congruent with inferences from previous genome-scale studies. It supports the major groupings of birds being Palaeognathae, Galloanseres and Neoaves~\citep{feng2020,stiller2024}. The analysis supports Neoaves as being split into 9 of the 10 major monophyletic lineages described previously~\citep{Jarvis2014-id}, and largely with maximal posterior supports (Fig.~\ref{fig:b10k}). The only exception is Afroaves, which is split into two groups as also suggested in several previous studies~\citep{Jarvis2014-id}. Importantly, this split is one of the few showing lower posterior support. Here, the lowest posterior support was found in nodes that are widely accepted to have occurred soon after the Cretaceous-Palaeogene (K-Pg) boundary and which pinpoint the uncertainty in the relationships among the 10 major groups of Neoaves. We also found low uncertainties primarily in the relative placements among Strisores, Aequornithes, Phaethontimorphae, Opisthocomiformes, and Cursorimorphae, which have long been recognised as difficult to place in the avian tree of life~\citep{reddy2017, houde2019}.

These results are consistent with the genome-wide disagreement in the relationships early after the post-K-Pg transition. Critically, the entropic likelihood Bayesian approach provides evidence that those nodes with low support cannot be placed confidently using nucleotide data alone. For illustration, fast molecular dating using penalised likelihood was performed as implemented in the \texttt{ape}~\citep{paradis2004} R package (function \textit{chronos}) using the avian maximum clade credibility tree and allowing for substantial variation in rates across lineages ($\Lambda$ = 0.001). We included 29 fossil constraints from the original study~\citep{stiller2024}, and a constraint on crown Neoaves to fall between 60 and 70 million years ago. As expected under these calibrations, the resulting dates support the so-called 'big-bang' scenario of rapid radiation scenario in most avian lineages occurring after the K-Pg mass extinction event~\citep{Brusatte2015-mo} and are overall in line with prior expectations.

\bigskip

\subsection{Conclusions}

In this paper, we have showcased the outstanding potential for distance-based approaches to perform highly efficient and accurate phylogenetic inference on large datasets while considering uncertainty in inferences, allowing for model selection, and incorporating complex models of evolution. Distance-based phylogenetics has traditionally been considered less effective than likelihood-based approaches due to the lack of marginalisation across internal nodes, the reduction of large amounts of sequence data into a single distance matrix, and the lack of model-based customisation. Yet, we demonstrate that these methods are robust to many of these perceived weaknesses. An inherent limitation of distance-based phylogenetics will be the inability to account for unknown states within a given tree to a greater extent than Felsenstein's likelihood-based approaches. Conversely, the expanding volume of phylogenetic data, in both site numbers and taxa, is placing substantial strain on maximum likelihood approaches, such that distance-based methods will play a critical role in genome-scale phylogenetic inference. This is already evident in major phylogenomics initiatives where computation times can extend into months, and where Bayesian solutions are unfeasible~\citep{feng2020}. Consequently, when considering large numbers of taxa (i.e., in the thousands) or site patterns (i.e., in the tens of millions), distance-based approaches are among the only viable solutions along with parsimony-based approaches.

Considering the inherent uncertainty in these distance-based phylogenetic inferences, the only viable approach until now has been bootstrapping, which provides limited meaningful information when the sample size of sites is very large. Our proposed entropic likelihood approach provides an intuitive alternative via Bayesian model-based inference as applied to distance phylogenetics. We have shown how distance-based posteriors produce sensible distributions of trees with close correspondence to the bootstrap. When factoring the quadratic complexity of distance-based approaches, the entropic likelihood also presents new avenues for Bayesian inference of large phylogenetic datasets, as we have shown on genome-scale data on avian taxa. 

The theory we introduce arrives at an approximation of Felsenstein's likelihood, but in which only inter-taxa pairs are considered. This entropic likelihood is very closely related to BME - by far the best-performing of all distance-based approaches. It has previously been shown that BME works as a special case of the weighted least-squares approach to phylogenetic inference~\citep{desper2004}. To show this, however, a fundamental assumption is that the variance is $\propto 2^\tau$, where $\tau$ is the topological distance taken, which is empirically justified but nevertheless a strong assumption. In contrast, we arrive at what is essentially BME with a distance matrix scaled by a constant, and show this is an approximation of the standard phylogenetic likelihood. Our theory therefore provides a different and robust justification for the use of BME and suggests that the standard BME objective function is approximately proportional to a likelihood. We prove that the error from this approximation is small, casting new confidence on fast methods for making accurate model-based inferences from genome-scale data.

\section{Data and Code availability}
All code relevant to reproducing the experiments is available online at \url{https://github.com/Neclow/bayesian_phylo_distances}~\citep{scheidwasser2024}. Instructions to access the publicly available datasets used in this study are included in the repository.

\section{Author contributions}
S.B, N.S, and M.J.P conceived the study. S.B supervised. S.B, N.S, and M.J.P designed the study. S.B, M.J.P, and N.S performed optimisation runs. S.B, M.J.P, and N.S performed analysis. S.B, M.J.P, and N.S drafted the first original draft. All authors contributed to editing the original draft. N.S and D.A.D contributed to revisions of the methodology. M.J.P, N.S, and S.B. drafted the appendix. N.S wrote the official implementation of Phylo2Vec.

\section{Competing interests}
The authors declare no competing interests.

\section{Funding}
M.J.P acknowledges support from his EPSRC DTP studentship, awarded by the University of Oxford to fund his DPhil in Statistics. D.A.D acknowledges support from the Novo Nordisk Foundation via the Data Science Emerging Investigator Award (NNF23OC0084647). S.B. acknowledges funding from the MRC Centre for Global Infectious Disease Analysis (reference MR/X020258/1), funded by the UK Medical Research Council (MRC). This UK funded award is carried out in the frame of the Global Health EDCTP3 Joint Undertaking.  S.B. is funded by the National Institute for Health and Care Research (NIHR) Health Protection Research Unit in Modelling and Health Economics, a partnership between UK Health Security Agency, Imperial College London and LSHTM (grant code NIHR200908). Disclaimer: “The views expressed are those of the author(s) and not necessarily those of the NIHR, UK Health Security Agency or the Department of Health and Social Care.” S.B. acknowledges support from the Novo Nordisk Foundation via The Novo Nordisk Young Investigator Award (NNF20OC0059309). S.B. acknowledges support from the Danish National Research Foundation via a chair grant (DNRF160) which also supports N.S. S.B. acknowledges support from The Eric and Wendy Schmidt Fund For Strategic Innovation via the Schmidt Polymath Award (G-22-63345). S.B and N.S acknowledge the Pioneer Centre for AI, DNRF grant number P1 as affiliate researchers. C.A.D receives support from the NIHR HPRU in Emerging and Zoonotic Infections, a partnership between the UK Health Security Agency, University of Liverpool, University of Oxford and Liverpool School of Tropical Medicine (grant code NIHR200907).



\bigskip\bigskip



\section{Figure captions}
\markboth{Figure captions}{Figure captions}
\begin{enumerate}
    \item \nameref{fig:entropy}
    \item \nameref{fig:error}
    \item \nameref{fig:entropic_vs_fels}
    \item \nameref{fig:KLdivergence}
    \item \nameref{fig:DS}
    \item \nameref{fig:b10k}
\end{enumerate}

\section{Table captions}
\markboth{Table captions}{Table captions}
\begin{enumerate}
    \item \nameref{tab:data}
    \item \nameref{tab:combined}
\end{enumerate}


\small
\begin{thebibliography}{51}
\providecommand{\natexlab}[1]{#1}
\providecommand{\selectlanguage}[1]{\relax}
\providecommand{\bibAnnoteFile}[1]{%
  \IfFileExists{#1}{\begin{quotation}\noindent\textsc{Key:} #1\\
  \textsc{Annotation:}\ \input{#1}\end{quotation}}{}}
\providecommand{\bibAnnote}[2]{%
  \begin{quotation}\noindent\textsc{Key:} #1\\
  \textsc{Annotation:}\ #2\end{quotation}}

\bibitem[{Bissiri et~al.(2016)Bissiri, Holmes, and Walker}]{Bissiri2016-nl}
Bissiri, P.~G., C.~C. Holmes, and S.~G. Walker. 2016. A general framework for
  updating belief distributions. J. R. Stat. Soc. Series B Stat. Methodol.
  78:1103--1130.
\input{Bissiri2016-nl}

\bibitem[{Bradbury et~al.(2018)Bradbury, Frostig, Hawkins, Johnson, Leary,
  Maclaurin, Necula, Paszke, Vander{P}las, Wanderman-{M}ilne, and
  Zhang}]{jax2018github}
Bradbury, J., R.~Frostig, P.~Hawkins, M.~J. Johnson, C.~Leary, D.~Maclaurin,
  G.~Necula, A.~Paszke, J.~Vander{P}las, S.~Wanderman-{M}ilne, and Q.~Zhang.
  2018. {JAX}: composable transformations of {P}ython+{N}um{P}y programs.
\input{jax2018github}

\bibitem[{Brusatte et~al.(2015)Brusatte, O'Connor, and
  Jarvis}]{Brusatte2015-mo}
Brusatte, S.~L., J.~K. O'Connor, and E.~D. Jarvis. 2015. The origin and
  diversification of birds. Curr. Biol. 25:R888--98.
\input{Brusatte2015-mo}

\bibitem[{Cavalli-Sforza and Edwards(1967)}]{Cavalli-Sforza1967-xh}
Cavalli-Sforza, L.~L. and A.~W. Edwards. 1967. Phylogenetic analysis. models
  and estimation procedures. Am. J. Hum. Genet. 19:233--257.
\input{Cavalli-Sforza1967-xh}

\bibitem[{De~Maio et~al.(2023)De~Maio, Kalaghatgi, Turakhia, Corbett-Detig,
  Minh, and Goldman}]{demaio2023}
De~Maio, N., P.~Kalaghatgi, Y.~Turakhia, R.~Corbett-Detig, B.~Q. Minh, and
  N.~Goldman. 2023. Maximum likelihood pandemic-scale phylogenetics. Nature
  Genetics Pages~1--7.
\input{demaio2023}

\bibitem[{Desper and Gascuel(2002)}]{Desper2002-uy}
Desper, R. and O.~Gascuel. 2002. Fast and accurate phylogeny reconstruction
  algorithms based on the minimum-evolution principle. J. Comput. Biol.
  9:687--705.
\input{Desper2002-uy}

\bibitem[{Desper and Gascuel(2004)}]{desper2004}
Desper, R. and O.~Gascuel. 2004. Theoretical foundation of the balanced minimum
  evolution method of phylogenetic inference and its relationship to weighted
  least-squares tree fitting. Mol. Biol. Evol. 21:587--598.
\input{desper2004}

\bibitem[{Efron et~al.(1996)Efron, Halloran, and Holmes}]{Efron1996-gv}
Efron, B., E.~Halloran, and S.~Holmes. 1996. Bootstrap confidence levels for
  phylogenetic trees. Proc. Natl. Acad. Sci. U. S. A. 93:13429--13434.
\input{Efron1996-gv}

\bibitem[{Felsenstein(1978)}]{Felsenstein1978-qc}
Felsenstein, J. 1978. The number of evolutionary trees. Syst. Biol. 27:27--33.
\input{Felsenstein1978-qc}

\bibitem[{Felsenstein(1981)}]{felsenstein1981}
Felsenstein, J. 1981. Evolutionary trees from dna sequences: a maximum
  likelihood approach. J. Mol. Evol. 17:368--376.
\input{felsenstein1981}

\bibitem[{Felsenstein(1983)}]{felsenstein1983}
Felsenstein, J. 1983. Statistical inference of phylogenies. J R Stat Soc Ser A
  146:246--262.
\input{felsenstein1983}

\bibitem[{Felsenstein(1985)}]{felsenstein1985}
Felsenstein, J. 1985. Confidence limits on phylogenies: an approach using the
  bootstrap. Evolution 39:783--791.
\input{felsenstein1985}

\bibitem[{Feng et~al.(2020)Feng, Stiller, Deng, Armstrong, Fang, Reeve, Xie,
  Chen, Guo, Faircloth et~al.}]{feng2020}
Feng, S., J.~Stiller, Y.~Deng, J.~Armstrong, Q.~Fang, A.~H. Reeve, D.~Xie,
  G.~Chen, C.~Guo, B.~C. Faircloth, et~al. 2020. Dense sampling of bird
  diversity increases power of comparative genomics. Nature 587:252--257.
\input{feng2020}

\bibitem[{Garey et~al.(1996)Garey, Near, Nonnemacher, and Nadler}]{garey1996}
Garey, J.~R., T.~J. Near, M.~R. Nonnemacher, and S.~A. Nadler. 1996. Molecular
  evidence for acanthocephala as a subtaxon of rotifera. J. Mol. Evol.
  43:287--292.
\input{garey1996}

\bibitem[{Guindon et~al.(2010)Guindon, Dufayard, Lefort, Anisimova, Hordijk,
  and Gascuel}]{Guindon2010-ip}
Guindon, S., J.-F. Dufayard, V.~Lefort, M.~Anisimova, W.~Hordijk, and
  O.~Gascuel. 2010. New algorithms and methods to estimate maximum-likelihood
  phylogenies: assessing the performance of {PhyML} 3.0. Syst. Biol.
  59:307--321.
\input{Guindon2010-ip}

\bibitem[{Hedges et~al.(1990)Hedges, Moberg, and Maxson}]{hedges1990}
Hedges, S.~B., K.~D. Moberg, and L.~R. Maxson. 1990. Tetrapod phylogeny
  inferred from 18s and 28s ribosomal rna sequences and a review of the
  evidence for amniote relationships. Mol. Biol. Evol. 7:607--633.
\input{hedges1990}

\bibitem[{Ho et~al.(2015)Ho, Duch{\^e}ne, and Duch{\^e}ne}]{Ho2015-up}
Ho, S. Y.~W., S.~Duch{\^e}ne, and D.~Duch{\^e}ne. 2015. Simulating and
  detecting autocorrelation of molecular evolutionary rates among lineages.
  Mol. Ecol. Resour. 15:688--696.
\input{Ho2015-up}

\bibitem[{Hoang et~al.(2018)Hoang, Chernomor, von Haeseler, Minh, and
  Vinh}]{Hoang2018-mi}
Hoang, D.~T., O.~Chernomor, A.~von Haeseler, B.~Q. Minh, and L.~S. Vinh. 2018.
  {UFBoot2}: Improving the ultrafast bootstrap approximation. Mol. Biol. Evol.
  35:518--522.
\input{Hoang2018-mi}

\bibitem[{Holmes(2003)}]{Holmes2003-cg}
Holmes, S. 2003. Bootstrapping phylogenetic trees: Theory and methods. Stat.
  Sci. 18:241--255.
\input{Holmes2003-cg}

\bibitem[{Houde et~al.(2019)Houde, Braun, Narula, Minjares, and
  Mirarab}]{houde2019}
Houde, P., E.~L. Braun, N.~Narula, U.~Minjares, and S.~Mirarab. 2019.
  Phylogenetic signal of indels and the neoavian radiation. Diversity 11.
\input{houde2019}

\bibitem[{Jarvis et~al.(2014)Jarvis, Mirarab, Aberer, Li, Houde, Li, Ho,
  Faircloth, Nabholz, Howard, Suh, Weber, da~Fonseca, Li, Zhang, Li, Zhou,
  Narula, Liu, Ganapathy, Boussau, Bayzid, Zavidovych, Subramanian,
  Gabald{\'o}n, Capella-Guti{\'e}rrez, Huerta-Cepas, Rekepalli, Munch,
  Schierup, Lindow, Warren, Ray, Green, Bruford, Zhan, Dixon, Li, Li, Huang,
  Derryberry, Bertelsen, Sheldon, Brumfield, Mello, Lovell, Wirthlin,
  Schneider, Prosdocimi, Samaniego, Vargas~Velazquez, Alfaro-N{\'u}{\~n}ez,
  Campos, Petersen, Sicheritz-Ponten, Pas, Bailey, Scofield, Bunce, Lambert,
  Zhou, Perelman, Driskell, Shapiro, Xiong, Zeng, Liu, Li, Liu, Wu, Xiao,
  Yinqi, Zheng, Zhang, Yang, Wang, Smeds, Rheindt, Braun, Fjeldsa, Orlando,
  Barker, J{\o}nsson, Johnson, Koepfli, O'Brien, Haussler, Ryder, Rahbek,
  Willerslev, Graves, Glenn, McCormack, Burt, Ellegren, Alstr{\"o}m, Edwards,
  Stamatakis, Mindell, Cracraft, Braun, Warnow, Jun, Gilbert, and
  Zhang}]{Jarvis2014-id}
Jarvis, E.~D., S.~Mirarab, A.~J. Aberer, B.~Li, P.~Houde, C.~Li, S.~Y.~W. Ho,
  B.~C. Faircloth, B.~Nabholz, J.~T. Howard, A.~Suh, C.~C. Weber, R.~R.
  da~Fonseca, J.~Li, F.~Zhang, H.~Li, L.~Zhou, N.~Narula, L.~Liu, G.~Ganapathy,
  B.~Boussau, M.~S. Bayzid, V.~Zavidovych, S.~Subramanian, T.~Gabald{\'o}n,
  S.~Capella-Guti{\'e}rrez, J.~Huerta-Cepas, B.~Rekepalli, K.~Munch,
  M.~Schierup, B.~Lindow, W.~C. Warren, D.~Ray, R.~E. Green, M.~W. Bruford,
  X.~Zhan, A.~Dixon, S.~Li, N.~Li, Y.~Huang, E.~P. Derryberry, M.~F. Bertelsen,
  F.~H. Sheldon, R.~T. Brumfield, C.~V. Mello, P.~V. Lovell, M.~Wirthlin,
  M.~P.~C. Schneider, F.~Prosdocimi, J.~A. Samaniego, A.~M. Vargas~Velazquez,
  A.~Alfaro-N{\'u}{\~n}ez, P.~F. Campos, B.~Petersen, T.~Sicheritz-Ponten,
  A.~Pas, T.~Bailey, P.~Scofield, M.~Bunce, D.~M. Lambert, Q.~Zhou,
  P.~Perelman, A.~C. Driskell, B.~Shapiro, Z.~Xiong, Y.~Zeng, S.~Liu, Z.~Li,
  B.~Liu, K.~Wu, J.~Xiao, X.~Yinqi, Q.~Zheng, Y.~Zhang, H.~Yang, J.~Wang,
  L.~Smeds, F.~E. Rheindt, M.~Braun, J.~Fjeldsa, L.~Orlando, F.~K. Barker,
  K.~A. J{\o}nsson, W.~Johnson, K.-P. Koepfli, S.~O'Brien, D.~Haussler, O.~A.
  Ryder, C.~Rahbek, E.~Willerslev, G.~R. Graves, T.~C. Glenn, J.~McCormack,
  D.~Burt, H.~Ellegren, P.~Alstr{\"o}m, S.~V. Edwards, A.~Stamatakis, D.~P.
  Mindell, J.~Cracraft, E.~L. Braun, T.~Warnow, W.~Jun, M.~T.~P. Gilbert, and
  G.~Zhang. 2014. Whole-genome analyses resolve early branches in the tree of
  life of modern birds. Science 346:1320--1331.
\input{Jarvis2014-id}

\bibitem[{Jukes et~al.(1969)Jukes, Cantor et~al.}]{jukes1969}
Jukes, T.~H., C.~R. Cantor, et~al. 1969. Evolution of protein molecules. Mamm.
  Protein Metab. 3:21--132.
\input{jukes1969}

\bibitem[{Klopfstein et~al.(2017)Klopfstein, Massingham, and
  Goldman}]{Klopfstein2017-uv}
Klopfstein, S., T.~Massingham, and N.~Goldman. 2017. More on the best
  evolutionary rate for phylogenetic analysis. Syst. Biol. 66:769--785.
\input{Klopfstein2017-uv}

\bibitem[{Kozlov et~al.(2019)Kozlov, Darriba, Flouri, Morel, and
  Stamatakis}]{kozlov2019}
Kozlov, A.~M., D.~Darriba, T.~Flouri, B.~Morel, and A.~Stamatakis. 2019.
  Raxml-ng: a fast, scalable and user-friendly tool for maximum likelihood
  phylogenetic inference. Bioinformatics 35:4453--4455.
\input{kozlov2019}

\bibitem[{Kumar(2022)}]{kumar2022}
Kumar, S. 2022. {Embracing Green Computing in Molecular Phylogenetics}.
  Molecular Biology and Evolution 39:msac043.
\input{kumar2022}

\bibitem[{Lefort et~al.(2015)Lefort, Desper, and Gascuel}]{lefort2015}
Lefort, V., R.~Desper, and O.~Gascuel. 2015. Fastme 2.0: a comprehensive,
  accurate, and fast distance-based phylogeny inference program. Mol. Biol.
  Evol. 32:2798--2800.
\input{lefort2015}

\bibitem[{Minh et~al.(2020)Minh, Schmidt, Chernomor, Schrempf, Woodhams,
  Von~Haeseler, and Lanfear}]{minh2020}
Minh, B.~Q., H.~A. Schmidt, O.~Chernomor, D.~Schrempf, M.~D. Woodhams,
  A.~Von~Haeseler, and R.~Lanfear. 2020. Iq-tree 2: new models and efficient
  methods for phylogenetic inference in the genomic era. Mol. Biol. Evol.
  37:1530--1534.
\input{minh2020}

\bibitem[{Morlon et~al.(2011)Morlon, Parsons, and Plotkin}]{Morlon2011-pe}
Morlon, H., T.~L. Parsons, and J.~B. Plotkin. 2011. Reconciling molecular
  phylogenies with the fossil record. Proc. Natl. Acad. Sci. U.S.A.
  108:16327--16332.
\input{Morlon2011-pe}

\bibitem[{Nguyen et~al.(2015)Nguyen, Schmidt, Von~Haeseler, and
  Minh}]{nguyen2015}
Nguyen, L.-T., H.~A. Schmidt, A.~Von~Haeseler, and B.~Q. Minh. 2015. Iq-tree: a
  fast and effective stochastic algorithm for estimating maximum-likelihood
  phylogenies. Mol. Biol. Evol. 32:268--274.
\input{nguyen2015}

\bibitem[{Paradis et~al.(2004)Paradis, Claude, and Strimmer}]{paradis2004}
Paradis, E., J.~Claude, and K.~Strimmer. 2004. {APE: analyses of phylogenetics
  and evolution in R language}. Bioinformatics 20:289--290.
\input{paradis2004}

\bibitem[{Paradis and Schliep(2019)}]{paradis2019}
Paradis, E. and K.~Schliep. 2019. ape 5.0: an environment for modern
  phylogenetics and evolutionary analyses in r. Bioinformatics 35:526--528.
\input{paradis2019}

\bibitem[{Parker and Ram(1999)}]{Parker1999-aa}
Parker, D.~S. and P.~Ram. 1999. The construction of huffman codes is a
  submodular (``convex'') optimization problem over a lattice of binary trees.
  SIAM J. Comput. 28:1875--1905.
\input{Parker1999-aa}

\bibitem[{Pauplin(2000)}]{pauplin2000}
Pauplin, Y. 2000. Direct calculation of a tree length using a distance matrix.
  J. Mol. Evol. 51:41--47.
\input{pauplin2000}

\bibitem[{Penn et~al.(2023)Penn, Scheidwasser, Penn, Donnelly, Duchêne, and
  Bhatt}]{penn2023leaping}
Penn, M.~J., N.~Scheidwasser, J.~Penn, C.~A. Donnelly, D.~A. Duchêne, and
  S.~Bhatt. 2023. {Leaping through Tree Space: Continuous Phylogenetic
  Inference for Rooted and Unrooted Trees}. Genome Biol. Evol. 15:evad213.
\input{penn2023leaping}

\bibitem[{Rannala and Yang(1996)}]{Rannala1996-zk}
Rannala, B. and Z.~Yang. 1996. Probability distribution of molecular
  evolutionary trees: a new method of phylogenetic inference. J. Mol. Evol.
  43:304--311.
\input{Rannala1996-zk}

\bibitem[{Rannala et~al.(2012)Rannala, Zhu, and Yang}]{rannala2012tail}
Rannala, B., T.~Zhu, and Z.~Yang. 2012. Tail paradox, partial identifiability,
  and influential priors in bayesian branch length inference. Molecular biology
  and evolution 29:325--335.
\input{rannala2012tail}

\bibitem[{Reddy et~al.(2017)Reddy, Kimball, Pandey, Hosner, Braun, Hackett,
  Han, Harshman, Huddleston, Kingston et~al.}]{reddy2017}
Reddy, S., R.~T. Kimball, A.~Pandey, P.~A. Hosner, M.~J. Braun, S.~J. Hackett,
  K.-L. Han, J.~Harshman, C.~J. Huddleston, S.~Kingston, et~al. 2017. Why do
  phylogenomic data sets yield conflicting trees? data type influences the
  avian tree of life more than taxon sampling. Syst. Biol. 66:857--879.
\input{reddy2017}

\bibitem[{Robinson and Foulds(1981)}]{robinson1981}
Robinson, D.~F. and L.~R. Foulds. 1981. Comparison of phylogenetic trees. Math.
  Biosci. 53:131--147.
\input{robinson1981}

\bibitem[{Roch(2006)}]{roch2006}
Roch, S. 2006. A short proof that phylogenetic tree reconstruction by maximum
  likelihood is hard. IEEE/ACM Trans. Comput. Biol. Bioinform. 3:92--94.
\input{roch2006}

\bibitem[{Scheidwasser et~al.(2024)Scheidwasser, Penn, Duchêne, and
  Bhatt}]{scheidwasser2024}
Scheidwasser, N., M.~J. Penn, D.~A. Duchêne, and S.~Bhatt. 2024. Bayesian
  distance-based phylogenetics for the genomics era.
\input{scheidwasser2024}

\bibitem[{Schliep(2011)}]{Schliep2011-zu}
Schliep, K.~P. 2011. phangorn: phylogenetic analysis in {R}. Bioinformatics
  27:592--593.
\input{Schliep2011-zu}

\bibitem[{Simion et~al.(2020)Simion, Delsuc, and Philippe}]{simion2020}
Simion, P., F.~Delsuc, and H.~Philippe. 2020. {To What Extent Current Limits of
  Phylogenomics Can Be Overcome?} Pages~2.1:1--2.1:34 \emph{in} {Phylogenetics
  in the Genomic Era} (C.~Scornavacca, F.~Delsuc, and N.~Galtier, eds.). {No
  commercial publisher | Authors open access book}.
\input{simion2020}

\bibitem[{Stadler(2011)}]{Stadler2011-dx}
Stadler, T. 2011. Simulating trees with a fixed number of extant species. Syst.
  Biol. 60:676--684.
\input{Stadler2011-dx}

\bibitem[{Stamatakis(2014)}]{stamatakis2014}
Stamatakis, A. 2014. {RAxML version 8: a tool for phylogenetic analysis and
  post-analysis of large phylogenies}. Bioinformatics 30:1312--1313.
\input{stamatakis2014}

\bibitem[{Stiller et~al.(2024)Stiller, Feng, Chowdhury, Rivas-Gonz{\'a}lez,
  Duch{\^e}ne, Fang, Deng, Kozlov, Stamatakis, Claramunt et~al.}]{stiller2024}
Stiller, J., S.~Feng, A.-A. Chowdhury, I.~Rivas-Gonz{\'a}lez, D.~A.
  Duch{\^e}ne, Q.~Fang, Y.~Deng, A.~Kozlov, A.~Stamatakis, S.~Claramunt, et~al.
  2024. Complexity of avian evolution revealed by family-level genomes. Nature
  629:851--860.
\input{stiller2024}

\bibitem[{Whidden et~al.(2020)Whidden, Claywell, Fisher, Magee, Fourment, and
  Matsen}]{Whidden2020-gp}
Whidden, C., B.~C. Claywell, T.~Fisher, A.~F. Magee, M.~Fourment, and F.~A.
  Matsen. 2020. Systematic exploration of the high likelihood set of
  phylogenetic tree topologies. Syst. Biol. 69:280--293.
\input{Whidden2020-gp}

\bibitem[{Whidden and Matsen~IV(2015)}]{Whidden2015-jn}
Whidden, C. and F.~A. Matsen~IV. 2015. Quantifying {MCMC} exploration of
  phylogenetic tree space. Syst. Biol. 64:472--491.
\input{Whidden2015-jn}

\bibitem[{Yang(2006)}]{yang2006}
Yang, Z. 2006. {Computational Molecular Evolution}. Oxford University Press.
\input{yang2006}

\bibitem[{Yang(2007)}]{Yang2007-ez}
Yang, Z. 2007. {PAML} 4: phylogenetic analysis by maximum likelihood. Mol.
  Biol. Evol. 24:1586--1591.
\input{Yang2007-ez}

\bibitem[{Yang and Rannala(2012)}]{yang2012}
Yang, Z. and B.~Rannala. 2012. Molecular phylogenetics: principles and
  practice. Nat. Rev. Genet. 13:303--314.
\input{yang2012}

\bibitem[{Yang and Yoder(2003)}]{yang2003}
Yang, Z. and A.~D. Yoder. 2003. Comparison of likelihood and Bayesian methods
  for estimating divergence times using multiple gene loci and calibration
  points, with application to a radiation of cute-looking mouse lemur species.
  Syst. Biol. 52:705--716.
\input{yang2003}

\end{thebibliography}

\normalsize


\normalsize
\clearpage
\setcounter{figure}{0}
\setcounter{table}{0}
\setcounter{algorithm}{0}
\makeatletter
\renewcommand{\thefigure}{S\@arabic\c@figure}
\renewcommand{\thetable}{S\@arabic\c@table}
\renewcommand{\thealgorithm}{S\@arabic\c@algorithm}
\makeatother

\markboth{Appendix}{Appendix}
\appendix
\section{Supplementary Information}
\subsection{Connection of entropic likelihood to BME}
In this section, we show that the percentage error in the BME approximation is small in the $\frac{K}{\theta} << 1$ limiting case.
\begin{theorem}
\label{th:bme}
    Define 
    \begin{equation}
    K = \sum_{a \neq b}\pi_a Q_{ab}
\end{equation}
Consider taking $Kt \to 0$ while keeping the stationary distribution $\pi$ and the ratios $\frac{Q_{ab}}{Q_{cd}}$ constant (i.e. one can change $t$ or scale the matrix $Q$ by a constant multiple). 

For some fixed $k_1$ and variable $k_2 > k_1$ define a linear BME approximation to the entropic likelihood $f(k_2)$ to be
\begin{equation}
    f(k_2) = \mathds{E}\bigg\{D_{ij}^S\bigg(\frac{k_1}{\theta}\bigg)\bigg\} + \frac{k_2 - k_1}{\theta}\bigg(\int_0^{\infty}S(s)\theta^2e^{-\theta s}ds\bigg)
\end{equation}
Then,
\begin{equation}
\label{eq:first_result}
    \frac{\mathds{E}\bigg\{D_{ij}^S\bigg(\frac{k_2}{\theta}\bigg)\bigg\} - \mathds{E}\bigg\{D_{ij}^S\bigg(\frac{k_1}{\theta}\bigg)\bigg\}}{f(k_2) - \mathds{E}\bigg\{D_{ij}^S\bigg(\frac{k_1}{\theta}\bigg)\bigg\}} = 1 + \mathcal{O}\left(\frac{1}{\log\bigg(\frac{K}{\theta}\bigg)}\right)
\end{equation}
and, moreover, the percentage error is small
\begin{equation}
    \frac{f(k_2) - \mathds{E}\bigg\{D_{ij}^S\bigg(\frac{k_2}{\theta}\bigg)\bigg\}}{f(k_2)} =\mathcal{O}\left(\frac{1}{\log\bigg(\frac{K}{\theta}\bigg)}\right)
\end{equation}
\end{theorem}
\textbf{Proof:} We begin with the true entropy difference between $\frac{k_1}{\theta}$ and $\frac{k_2}{\theta}$, which is
\begin{equation}
    \mathds{E}\bigg\{D_{ij}^S\bigg(\frac{k_2}{\theta}\bigg)\bigg\} - \mathds{E}\bigg\{D_{ij}^S\bigg(\frac{k_1}{\theta}\bigg)\bigg\} = \int_{\frac{k_1}{\theta}}^{\frac{k_2}{\theta}}\bigg\{e^{-\theta t}(S'(t) + \theta S(t)) + \int_0^{t}S(s)\theta^2e^{-\theta s}ds\bigg\} dt
\end{equation}
This can be rewritten as
\begin{equation}
     \mathds{E}\bigg\{D_{ij}^S\bigg(\frac{k_2}{\theta}\bigg)\bigg\} - \mathds{E}\bigg\{D_{ij}^S\bigg(\frac{k_1}{\theta}\bigg)\bigg\} = \frac{k_2-k_1}{\theta}\bigg(\int_0^{\infty}S(s)\theta^2e^{-\theta s}ds\bigg) + \int_{\frac{k_1}{\theta}}^{\frac{k_2}{\theta}}\bigg\{e^{-\theta t}(S'(t) + \theta S(t)) -\int_t^{\infty}S(s)\theta^2e^{-\theta s}ds\bigg\} dt
\end{equation}
which, after twice integrating $\int_t^{\infty}S(s)\theta^2e^{-\theta s}ds$ by parts (integrating the exponential and differentiating the entropy), is
\begin{align}
     \mathds{E}\bigg\{D_{ij}^S\bigg(\frac{k_2}{\theta}\bigg)\bigg\} - \mathds{E}\bigg\{D_{ij}^S\bigg(\frac{k_1}{\theta}\bigg)\bigg\} &= \frac{k_2-k_1}{\theta}\bigg(\int_0^{\infty}S(s)\theta^2e^{-\theta s}ds\bigg) +  \int_{\frac{k_1}{\theta}}^{\frac{k_2}{\theta}}\int_t^{\infty}S''(s)e^{-\theta s}ds dt \\
     &= f(k_2) - \mathds{E}\bigg\{D_{ij}^S\bigg(\frac{k_1}{\theta}\bigg)\bigg\} +  \int_{\frac{k_1}{\theta}}^{\frac{k_2}{\theta}}\int_t^{\infty}S''(s)e^{-\theta s}ds dt
\end{align}
Thus, 
\begin{equation}
    \frac{\mathds{E}\bigg\{D_{ij}^S\bigg(\frac{k_2}{\theta}\bigg)\bigg\} - \mathds{E}\bigg\{D_{ij}^S\bigg(\frac{k_1}{\theta}\bigg)\bigg\}}{f(k_2) - \mathds{E}\bigg\{D_{ij}^S\bigg(\frac{k_1}{\theta}\bigg)\bigg\}} = 1 + \frac{\int_{\frac{k_1}{\theta}}^{\frac{k_2}{\theta}}\int_t^{\infty}S''(s)e^{-\theta s}ds dt}{\frac{k_2-k_1}{\theta} \times \theta^2 \int_0^{\infty}S(s)e^{-\theta s}ds}
\end{equation}
and so we simply need to justify that
\begin{equation}
\label{eq:bigfrac}
    \frac{\int_{\frac{k_1}{\theta}}^{\frac{k_2}{\theta}}\int_t^{\infty}S''(s)e^{-\theta s}ds dt}{\frac{k_2-k_1}{\theta} \times \theta^2 \int_0^{\infty}S(s)e^{-\theta s}ds} = \mathcal{O}\left(\frac{1}{\log\bigg(\frac{K}{\theta}\bigg)}\right)
\end{equation}
To begin, we consider the integral in the denominator. Note that
\begin{equation}
    \int_0^{\infty}\theta S(s)e^{-\theta s}ds = \int_0^{\infty}S\bigg(\frac{\tau}{\theta}\bigg)e^{-\tau}d\tau
\end{equation}
Now, choose some $C$, and note that (as $S(t)$ must be bounded between $0$ and above by $-\log(N_s)$, where $N_s$ is the number of states)
\begin{equation}
    \bigg|\int_C^{\infty}S\bigg(\frac{\tau}{\theta}\bigg)e^{-\tau}d\tau\bigg| \leq -\log(N_s)e^{-C}
\end{equation}
and so, provided $C >> 1$, this is exponentially small.

Moreover, in the region $\frac{KC}{\theta} << 1$, we can use Lemma~\ref{lem:Better0} to approximate
\begin{equation}
\label{eq:simpleint}
     \int_0^{C}S(\frac{\tau}{\theta})e^{-\tau}d\tau = \int_0^{C}\bigg\{\frac{K \tau}{\theta}\log\bigg(\frac{K\tau}{\theta}\bigg) + \mathcal{O}\bigg(\frac{K\tau}{\theta}\bigg)\bigg\}e^{-\tau}d\tau
\end{equation}
The two conditions on $C$ can be simultaneously satisfied by, for example, setting $C = \bigg(\frac{K}{\theta}\bigg)^{-0.5}$. 

Now, note that
\begin{equation}
     \int_0^{C}\mathcal{O}\bigg(\frac{K\tau}{\theta}\bigg)e^{-\tau}d\tau = \mathcal{O}\bigg(\frac{K}{\theta}\bigg)\int_0^{C}\tau e^{-\tau}d\tau = \mathcal{O}\bigg(\frac{K}{\theta}\bigg)
\end{equation}
and hence, the leading order term in~\ref{eq:simpleint} is
\begin{equation}
\label{eq:denominator}
     \int_0^{\infty}\theta S(s)e^{-\theta s}ds \sim \frac{K \tau}{\theta}\log\bigg(\frac{K\tau}{\theta}\bigg)\int_0^{C}e^{-\tau}d\tau \sim  \frac{K \tau}{\theta}\log\bigg(\frac{K\tau}{\theta}\bigg)
\end{equation}
again using the fact that $e^{-C}$ is exponentially small. 

Now, to simplify the numerator of~\ref{eq:bigfrac}, we begin by considering the inner integral evaluated at some possible value of $t$, which we set to be $\frac{k}{\theta}$ for $k \in (k_1,k_2)$ (and therefore $k = \mathcal{O}(1)$). Now,
\begin{equation}
    \int_{\frac{k}{\theta}}^{\infty}\theta S''(s)e^{-\theta s}ds = \int_{k}^{\infty}S''\bigg(\frac{\tau}{\theta}\bigg)e^{-\tau} d\tau
\end{equation}
Again, we can solve this problem by splitting the integral. Note that, using Lemma~\ref{lem:largeSdd}, $S''(s) \to 0$ as $s \to \infty$, so it must be bounded by some $\mathcal{S}$ in the region $[k,\infty)$. Defining
\begin{equation}
    X = \bigg(\frac{K}{\theta}\bigg)^{-0.5}
\end{equation}
we therefore see that
\begin{equation}
    \bigg|\int_{X}^{\infty}S''\bigg(\frac{\tau}{\theta}\bigg)e^{-\tau} d\tau\bigg| \leq \mathcal{S}e^{-X}
\end{equation}
which is exponentially small as $X \to 0$. Moreover, in the region $\tau << X$, we have $\frac{K\tau}{\theta} << 1$ and so, using Lemma~\ref{lem:Better0}, $S''(t) \sim \frac{K}{t}$. Thus,
\begin{equation}
    \bigg|\int_{k}^{X}S''\bigg(\frac{\tau}{\theta}\bigg)e^{-\tau} d\tau\bigg| \sim \bigg| \int_{k}^{X}\frac{K\theta}{\tau}e^{-\tau}d\tau\bigg| \leq \bigg|\frac{K \theta}{k_1} \int_{0}^{\infty}e^{-\tau}d\tau\bigg| =  \frac{K \theta}{k_1}
\end{equation}
Therefore, using these results in~\ref{eq:bigfrac}, and noting that we have multiplied each of the integrals by $\theta$ in their consideration (which of course cancels)
\begin{equation}
    \left|\frac{\int_{\frac{k_1}{\theta}}^{\frac{k_2}{\theta}}\int_t^{\infty}S''(s)e^{-\theta s}ds dt}{\frac{k_2-k_1}{\theta} \times \theta^2 \int_0^{\infty}S(s)e^{-\theta s}ds}\right| \lesssim \left|\frac{\int_{\frac{k_1}{\theta}}^{\frac{k_2}{\theta}}\frac{K \theta}{k_1}dt}{\frac{k_2-k_1}{\theta} \times\theta^2 \frac{K}{\theta}\log\bigg(\frac{K}{\theta}\bigg)}\right| = \frac{1}{k_1\log\bigg(\frac{K}{\theta}\bigg)} 
\end{equation}
as required for equation \ref{eq:first_result}. It is worth noting that, while $k_1$ is fixed throughout this theorem, the result would not hold in a $k_1 \to 0$ limit as the approximation breaks down for extremely small branch lengths.

The final result follows as rearranging \ref{eq:first_result} shows that
\begin{equation}
     \frac{f(k_2) - \mathds{E}\bigg\{D_{ij}^S\bigg(\frac{k_2}{\theta}\bigg)\bigg\}}{f(k_2)} = \left(\frac{\mathbb{E}\bigg\{D_{ij}^S\bigg(\frac{k_1}{\theta}\bigg)\bigg\}}{f(k_2)}-1\right)\mathcal{O}\left(\frac{1}{\log\bigg(\frac{K}{\theta}\bigg)}\right)
\end{equation}
and so, using the fact that the linear approximation is monotonic, $|f(k_2)| > \bigg|\mathbb{E}\bigg\{D_{ij}^S\bigg(\frac{k_1}{\theta}\bigg)\bigg\}\bigg|$ and the result follows.

\subsection{Supporting Lemmas}
\subsubsection{Independence of simulation root}

We claim that the distribution of sites on the tree is independent of the choice of simulation root. Define $p(g,r)$ to be the probability of the site pattern $g$ given that the simulation root is $r$.

\begin{lemma}
\label{lem:sim}
    If $\mathcal{N}$ is the set of nodes in $\mathcal{T}$, and our site evolution CTMC is reversible, then
    \begin{equation}
        p(g,r) = p(g,s) \quad \forall r,s \in \mathcal{N}
    \end{equation}
\end{lemma}
\textbf{Proof:} As before, define the set of edges to be $\mathcal{E} = \{(e_i^1,e_i^2)| i =0,1,..\}$, where $e_i^1$ and $e_i^2$ are the nodes which this edge connects, such that $e^1_i$ is the closer to the simulation root. Suppose that $z_i^j$ is the site value on node $e_i^j$ and that $b_i$ is the length of edge $(e_i^1,e_i^2)$.

Suppose first that $r$ and $s$ are adjacent so that they share an edge. Suppose that when the root is $r$, this edge is labelled as $e_1$ in $E$. Then, note that, using detailed balance,
\begin{equation}
    \pi_{z_r} P_{z_1^1,z_1^2}(t_i) =  \pi_{z_r} P_{z_r,z_s}(t_i) = \pi_{z_s} P_{z_s,z_r}(t_i) 
\end{equation}
When the root is $s$, it is also possible to label this edge as $e_1$, but now $z_1^1 = s$ and $z_1^2 = r$. Thus,
\begin{equation}
    p(g,s) = \pi_{z_s} P_{z_1^2,z_1^1}(t_i) \prod_{i\geq 2}P_{z_i^1,z_i^2}(t_i) = \pi_{z_r} P_{z_r,z_s}(t_i) \prod_{i\geq 2}P_{z_i^1,z_i^2}(t_i) = p(g,r)
\end{equation}
and so the likelihood is unchanged.

For any pair of non-adjacent nodes, $r$ and $s$, one can find the path of nodes $y_1,..,y_m$ between them. Then, $p(g,r) = p(g,s)$ follows from the fact that $p(g,r) = p(g,y_1) = p(g,y_2)... = p(g,s)$.

\subsubsection{Pairwise distributions}
\begin{lemma}
\label{lem:pairs}
    Let $M_t$ be a stationary reversible substitution CTMC. For a given pair of taxa, $x$ and $y$, which are distance $t$ apart in a (known) tree, $\mathcal{T}$, define $V_x$ and $V_y$ to be the values of a particular state in those two taxa. Then,
\begin{equation}
    (M_t,M_0) \stackrel{d}{=} (V_x,V_y)
\end{equation} 
\end{lemma}
\textbf{Proof:} Considering $r$ as a root, one can define the \textit{most recent common ancestor}, $m$, of $x$ and $y$ to be the node furthest from $r$ that is on both the path between $x$ and $r$ and the path between $y$ and $r$. (Note that it is possible that $m = r$).

Define $\mathcal{V}$ to be the set of possible states for a node, and define $\pi_s$ to be the stationary distribution. Then, using $V_r$ and $V_m$ to be the state at $r$ and $m$ respectively,
\begin{align}
    \mathbb{P}(V_x = a, V_y = b) &= \sum_{c\in \mathcal{V}} \mathbb{P}(V_x = a, V_y = b | V_r = c) \mathbb{P}(V_r = c)\\
    &= \sum_{c \in \mathcal{V}} \mathbb{P}(V_x = a, V_y = b | V_r = c) \pi_c\\
    &= \sum_{c,d \in \mathcal{V}}\mathbb{P}(V_x = a, V_y = b | V_m = d, V_r = c) \mathbb{P}(V_m=d | V_r = c) \pi_c\\
    &= \sum_{c,d \in \mathcal{V}}\mathbb{P}(V_x = a, V_y = b | V_m = d) \mathbb{P}(V_m=d | V_r = c) \pi_c
\end{align}
Now, note that (defining $t_r$ to be the distance between $r$ and $m$), using the detailed balance equations gives
\begin{equation}
    \sum_{c \in \mathcal{V}}\mathbb{P}(V_m=d | V_r = c) \pi_c =  \sum_{c \in \mathcal{V}}P_{cd}(t_r)\pi_c = \sum_{c \in \mathcal{V}}P_{dc}(t_r)\pi_d = \pi_d 
\end{equation}
and so,
\begin{equation}
    \mathbb{P}(V_x = a, V_y = b) = \sum_{d \in \mathcal{V}}\mathbb{P}(V_x = a, V_y = b | V_m = d)\pi_d
\end{equation}
Now, the substitution process on the path from $m$ to $a$ is independent of the substitution process on the path from $m$ to $b$. Thus, if the distance from $m$ to $a$ is $t_a$ and the distance from $m$ to $b$ is $t_b$, 
\begin{equation}
     \mathbb{P}(V_x = a, V_y = b) = \sum_{d \in \mathcal{V}}\pi_d P_{da}(t_a)P_{db}(t_b)
\end{equation}
Using detailed balance shows
\begin{equation}
   \mathbb{P}(V_x = a, V_y = b) = \sum_{d \in \mathcal{V}}\pi_a P_{ad}(t_a)P_{db}(t_b)
\end{equation}
Finally, note that
\begin{align}
   \pi_a \sum_{d \in \mathcal{V}} P_{ad}(t_a)P_{db}(t_b) &= \mathbb{P}(M_0 = a)\sum_{d \in \mathcal{V}}\mathbb{P}(M_{t_a} = d | M_{0} = a)\mathbb{P}(M_{t_a + t_b} = b | M_{t_a} = d) \\
    &= \mathbb{P}(M_0 = a)\mathbb{P}(M_{t_a + t_b} = b | M_0 = a) \\
    &= \mathbb{P}(M_0 = a, M_{t_a + t_b} = b)
\end{align}
and the result follows from the fact that $t_a + t_b = t$.

\subsubsection{Entropic distance for a general branching process}
Suppose that a tree $\mathcal{T}$ is generated as follows. Begin by choosing a total tree time $T$, and start the tree at a root node $\rho$ (which can ultimately be removed from the tree to make it unrooted). We assume that two iid trees are connected by $\rho$. For each of these trees, we simulate a branch of random length $T_0$ according to some pdf $f$, with cumulative distribution function (cdf) $F$.  If $T > T_0$, then we create a leaf node at a distance $T$ from the root and terminate the generation of this tree. Otherwise, if $T_0 < T$, then we create an internal node at a distance $T_0$ from the root. We then generate two independent trees rooted at this internal node, but with a total time of $T-T_0$. The construction process stops when there are no more trees that need to be generated.
\begin{lemma}
\label{lem:branchgen}
    Given two taxa $i$ and $j$ which are a distance of $2\tau$ apart on $\mathcal{T}$, the expected entropic distance, $\mathbb{E}(D^S_{ij})$ satisfies
    \begin{equation}
        \mathbb{E}(D^S_{ij}) = 2\mathcal{L}^{-1}\bigg(\frac{\bar{S}(p) + (p-1)\bar{g}(p)}{1 - \bar{f}(p) }\bigg)\bigg|_{\tau}
    \end{equation}
    where $\bar{g}(p)$ is the Laplace transform of $S(\tau)F(\tau)$.
\end{lemma}
\textbf{Proof:} Define $a$ to be the most recent common ancestor of $i$ and $j$ (that is, the node furthest from $x_0$ which is on the path from both $i$ to $x_0$ and $j$ to $x_0$). As the total distance from each leaf node to the root is fixed, it is necessary that both the distance from $a$ to $i$ and the distance from $a$ to $j$ is $\frac{\tau}{2}$.

The two subtrees rooted at $a$ are therefore iid, and so $\mathbb{E}(D^S_{ij}) = \frac{\mathbb{E}(D^S_{ia})}{2}$. We can calculate $\mathbb{E}(D^S_{ia})$ using the self-similarity property of the tree-generation process. Conditioning on the first branch length, $T_0$, of this tree
\begin{equation}
   h(\tau):=  \mathbb{E}(D^S_{ia}) = \int_0^{\infty} \mathbb{E}(D^S_{ia}| T_0 = t)f(t)dt 
\end{equation}
If $T_0 > \tau$, then this subtree has only one branch, and therefore $D^S_{ia} = S(\tau)$. Otherwise, we get an entropy of $S(T_0)$, and then get the entropy of a subtree of length $S(\tau-T_0)$. Thus,
\begin{equation}
\label{eq:ode}
    h(\tau) = (1-F(\tau))S(\tau) + \int_0^{\tau}(h(\tau - t) + S(t))f(t)dt
\end{equation}
To Laplace transform this equation, it is helpful to note that, as $\int_0^{\tau}S(t)f(t)dt$ is the integral of $S(\tau)f(\tau)$, its Laplace transform $\bar{G}(p)$ satisfies
\begin{equation}
    \bar{g}(p) = p\bar{G}(p) - G(0) = p\bar{G}(p)
\end{equation}
and hence, Laplace transforming the equation shows
\begin{equation}
    \bar{h}(p) = \bar{S}(p) - \bar{g}(p) + \bar{h}(p)\bar{f}(p) + \frac{1}{p}g(p)
\end{equation}
and hence
\begin{equation}
    h(\tau)= \mathcal{L}^{-1}\bigg(\frac{\bar{S}(p) + (\frac{1}{p}-1)\bar{g}(p)}{1 - \bar{f}(p) }\bigg)\bigg|_{\tau}
\end{equation}
as required.
\subsubsection{Entropic distance for a Markovian branching process}
\begin{lemma}
    \label{lem:branchexp}
    With the setup of Lemma~\ref{lem:branchexp}, if the branch lengths are exponentially distributed with mean $\frac{1}{\theta}$, then
    \begin{equation}
        \mathbb{E}(D^S_{ij}) =2 \bigg(\int_0^{\tau}\bigg[e^{-\theta t}(S'(t) + \theta S(t)) + \int_0^{t}S(s)\theta^2 e^{-\theta s}ds\bigg] dt\bigg)
    \end{equation}
\end{lemma}
\textbf{Proof:} In this case, (\ref{eq:ode}) can be rewritten as
\begin{align}
    h(\tau) &= e^{-\theta \tau} S(\tau) + \int_0^{\tau}(h(\tau - t) + S(t))\theta e^{-\theta t} dt\\
    &= e^{-\theta \tau} S(\tau) + \int_0^{\tau}h(t)\theta e^{-\theta (\tau - t)} dt + \int_0^{\tau}S(t)\theta e^{-\theta t}dt
\end{align}
and hence
\begin{equation}
     h(\tau)e^{\theta \tau} = S(\tau) + \int_0^{\tau}h(t)\theta e^{\theta (t)} dt + e^{\theta \tau}\int_0^{\tau}S(t)\theta e^{-\theta t}dt
\end{equation}
Differentiating for $\tau > 0$ gives
\begin{equation}
    e^{\theta\tau}(h'(\tau) + \theta h(\tau)) = S'(\tau) + h(\tau)\theta e^{\theta \tau} + \theta S(\tau) + \theta e^{\theta \tau}\int_0^{\tau}S(t)\theta e^{-\theta t}dt
\end{equation}
Hence, 
\begin{equation}
    h'(\tau) = e^{-\theta \tau}(S'(\tau) + \theta S(\tau)) + \int_0^{\tau}\theta^2S(t)e^{-\theta t}dt
\end{equation}
and so, using Lemma~\ref{lem:Better0} to show that the singularity of $S'(t)$ at $t=0$ is logarithmic, and therefore integrable
\begin{equation}
    h(\tau) = \int_0^{\tau}\bigg[e^{-\theta t}(S'(t) + \theta S(t)) + \int_0^{t}S(s)\theta^2 e^{-\theta s}ds\bigg] dt
\end{equation}
\subsubsection{Maximum likelihood estimation of \texorpdfstring{$\theta$}{θ} for a Markovian branching process}
\begin{lemma}
\label{lem:theta}
    Given the branching process in Lemma~\ref{lem:branchexp} and a total tree length of $\mathcal{L}$ the maximum likelihood estimator $\hat{\theta}$ of $\theta$ is
    \begin{equation}
        \hat{\theta} = \frac{n-2}{\mathcal{L}}
    \end{equation}
\end{lemma}
\textbf{Proof:} We can construct our tree using a ``time-to-next-event'' construction. That is, we consider the events $\mathcal{E}_1,\mathcal{E}_2,...$ where a branch terminates, or when the time reaches $T_0$ and the process ends, in increasing order of the time at which they terminate. We use $l_k$ to denote the time at which $\mathcal{E}_k$ occurs.

Using the memoryless property of the exponential distribution, immediately after $\mathcal{E}_k$, we have $k+2$ active branches and hence
\begin{equation}
    l_{k+1} \sim l_{k} + \min\bigg(T_0-l_k,\text{Exp}((k+2)\theta)\bigg)
\end{equation}

We know that, as we have $n$ leaf nodes, there must have been $n-2$ splitting events up to time $T_0$. For a given set of times $l_k$ at which these events happened, we have a probability density function (pdf), $f(\boldsymbol{l})$
\begin{equation}
    f(\boldsymbol{l}) = \bigg[\prod_{i = 1}^{n-2}(i+1)\theta e^{-(i+1)\theta(l_i-l_{i-1})}\bigg]e^{-n\theta(T_0-l_{n-2})}
\end{equation}
This is proportional to
\begin{equation}
    g(\boldsymbol{l}) = \theta^{n-2}\exp\bigg[-\sum_{i=1}^{n-2}(i+1)\theta(l_i-l_{i-1}) -n\theta(T_0-l_{n-2})\bigg]
\end{equation}
Now, we know that the total length of the tree is
\begin{equation}
    \mathcal{L} = \sum_{i=1}^{n-2}(i+1)(l_i - l_{i-1}) + (T_0 -l_{n-2})n
\end{equation}
by summing the $(i+1)$ branch lengths between events $\mathcal{E}_{i}$ and $\mathcal{E}_{i-1}$. Thus,
\begin{equation}
    g(\boldsymbol{l}) = \theta^{n-2}\exp\bigg[-\theta \mathcal{L}\bigg]
\end{equation}
We can find the maximiser of this by noting
\begin{equation}
    \log(g) = (n-2)\log(\theta) - \mathcal{L}\theta
\end{equation}
and hence, after differentiating
\begin{equation}
    \hat{\theta}  = \frac{n-2}{\mathcal{L}}
\end{equation}
as required.
\subsubsection{Varying the sampling distribution}
\begin{lemma}
\label{lem:k2}
    \begin{equation}
   H(X,Y|\mathcal{T},\boldsymbol{\beta}) \approx  \kappa_2H(X,Y|\mathcal{U},\boldsymbol{b}^*) 
\end{equation}
for some $\kappa_2 \in [1,2]$
\end{lemma}
\textbf{``Proof'':} We do not prove this rigorously, but instead provide a rough justification as to why we expect this result to hold.

We define $p_i$ to be the probability of sequence $i$ in the true tree $\mathcal{T}$ and $q_i$ to be the probability in $\mathcal{U}$

Rewriting the equation stated in the lemma in these terms, we seek to show
\begin{equation}
    \sum_i p_i\log(q_i) \approx \kappa_2 \sum_i q_i \log(q_i)
\end{equation}

Now, consider the solution to
\begin{equation}
    \max_{\boldsymbol{q}}\{\sum_i p_i \log(q_i) \bigg| \sum_i q_i\log(q_i) = C , \sum_i q_i = 1\}
\end{equation}
When $\mathcal{U}$ is close to $\mathcal{T}$, we know that finding the balanced minimum evolution tree lengths will make $q_i$ approximate $p_i$, and should therefore approximately maximise $\sum_i p_i \log(q_i)$, given that the entropy of $\sum_i q_i\log(q_i)$ has increased to some constant $C$. This approximation is not perfect, and so $\boldsymbol{q}$ will not exactly solve this optimization problem, but it is a good estimate with which we can examine the effect on our entropies. To solve this problem we consider a Lagrange multiplier to get
\begin{equation}
    \mathcal{L}(\boldsymbol{q},\lambda,\mu) = \sum_i (p_i-\lambda q_i) \log(q_i) - \mu q_i
\end{equation}
Differentiating and setting to zero gives
\begin{equation}
    \frac{p_i}{q_i} - \lambda(1 + \log(q_i)) = \mu
\end{equation}
Multiplying by $q_i$
\begin{equation}
    p_i - \lambda q_i - \lambda q_i \log(q_i) - \mu q_i = 0
\end{equation}
and summing over $i$ gives an equation
\begin{equation}
    1 - \lambda - \lambda C - \mu = 0 
\end{equation}
and so
\begin{equation}
    \mu = 1 - (C+1)\lambda
\end{equation}
and hence
\begin{equation}
    p_i - \lambda q_i - \lambda q_i \log(q_i) - (1-(C+1)\lambda)q_i = 0
\end{equation}
so
\begin{equation}
    p_i = \lambda q_i \log(q_i) + (1-C\lambda)q_i= q_i(\lambda \log(q_i) + (1-C\lambda))
\end{equation}
From here, we work in the case where $|\mathcal{G}|$ is large and suppose that $C = \phi \log(\frac{1}{|\mathcal{G}|})$ for $\phi \in (0,1)$ and $\phi = \mathcal{O}(1)$ (noting that the maximal entropy is $\log(\frac{1}{|\mathcal{G}|})$. Thus, $|C|>>1$ also.

We know that for each $i$, as $p_i \geq 0$
\begin{equation}
    q_i \geq \exp\bigg[C - \frac{1}{\lambda}\bigg]
\end{equation}
If this held to equality, summing over $i$ would yield
\begin{equation}
    \exp\bigg[C - \frac{1}{\lambda}\bigg] = \frac{1}{|\mathcal{G}|}
\end{equation}
We therefore instead suppose that there exists some $\gamma = \mathcal{O}(1)$ such that
\begin{equation}
    \exp\bigg[C - \frac{1}{\lambda}\bigg] = \bigg(\frac{1}{|\mathcal{G}|}\bigg)^{\gamma}
\end{equation}
which means
\begin{equation}
    \lambda = \frac{1}{C - \gamma \log(\frac{1}{|\mathcal{G}|})} = \frac{1}{C(1-\gamma \phi)}
\end{equation}

Now, multiplying our equation by $\frac{p_i}{q_i}$, and summing over $i$, we have
\begin{equation}
    \sum_i \frac{p^2_i}{q_i} = \lambda \sum_i p_i \log(q_i) + (1-C\lambda)
\end{equation}
We suppose that our trees are close so that $\sum_i \frac{p^2_i}{q_i} = \mathcal{O}(1)$ (which follows as $\frac{p_i}{q_i} \approx 1$ and hence we are left with approximately $\sum_i p_i = 1$). Ignoring this term then yields
\begin{equation}
     \sum_ip_i \log(q_i)  \approx C + \frac{1}{\lambda} = C(2-\gamma \phi)
\end{equation}
Note that the case $\gamma = \phi = 1$ means that $C$ is the maximal entropy for this distribution and therefore all the $q_i$ are equal and so, as expected, we would get $ \sum_ip_i \log(q_i) = C$. However, in general, we expect $\gamma \phi \in (0,1)$ (noting that, necessarily $|\sum_i p_i\log(q_i)| \geq |\sum_i p_i \log(p_i)|$.

This is the required result, as the left-hand-side (under this approximate construction) is equal to $H(X,Y|\mathcal{U},\boldsymbol{b}^*)$ and the right-hand-side equal to $H(X,Y|\mathcal{T},\boldsymbol{\beta})$. Hence, we expect that 
\begin{equation}
   \kappa_2 = C(2-\gamma \phi) \in (1,2)
\end{equation}
Of course, $\gamma$ and $\phi$ are functions of $C$, but provided we only make small changes to $C$, we expect them to vary slowly, and so the required linear formula will approximately hold for close trees.
\subsubsection{Large time behaviour of  \texorpdfstring{$S''(t)$}{S''(t)}}
\begin{lemma}
\label{lem:largeSdd}
\begin{equation}
    \lim_{t \to \infty}(S''(t)) = 0
\end{equation}

\end{lemma}
\textbf{Proof:} First, note that, as the substitution CTMC is reversible,
\begin{equation}
    \pi_iQ_{ij} = \pi_jQ_{ji} \Rightarrow \sqrt{\pi_i}Q_{ij}\frac{1}{\sqrt{\pi_j}} = \frac{1}{\sqrt{\pi_i}}Q_{ji} \sqrt{\pi_j}
\end{equation}
Hence, the matrix $\tilde{Q}$ given by
\begin{equation}
    \tilde{Q}_{ij} =\sqrt{\pi_i}Q_{ij}\frac{1}{\sqrt{\pi_j}}
\end{equation}
is symmetric and therefore diagonalisable by spectral theorem. Noting that
\begin{equation}
    \tilde{Q} = \Pi^{\frac{1}{2}}Q\Pi^{-\frac{1}{2}}
\end{equation}
where $\Pi$ is a diagonal matrix with entries $\pi_i$, $\tilde{Q}$ is similar to $Q$, the matrix $Q$ is therefore diagonalisable. Now, applying this to the forward equations gives
\begin{equation}
    P(t) = \exp(Qt) = M^{-1}e^{Dt}MP(0)
\end{equation}
for some matrix $M$ and a diagonal matrix $D$. Hence, $P(t)$ is a weighted sum of exponentials, and its asymptotic behaviour is controlled by the leading order term. As, by the ergodic theorem, $P$ has a finite limit,
\begin{equation}
    \lim_{t \to \infty}P_{ab}(t) = \pi_b,
\end{equation}
the dominant exponential terms must have a non-positive exponent, meaning
\begin{equation}
    P^{(k)}_{ij}(t) \to 0 \quad \forall k \geq 1
\end{equation}
where $P^{(k)}_{ij}(t)$ denotes the $k^{\text{th}}$ derivative of $P_{ij}(t)$.

Recall that
\begin{equation}
    -S(t) = \sum_{a,b}\pi_aP_{ab}(t)\log(P_{ab}(t))
\end{equation}
One could show the result of the lemma by computing the second derivative manually - instead, we use a briefer, though less detailed argument. Note that $S''(t)$ will be a linear function of each $P''_{ab}$ and a quadratic function of each $P'_{ab}$. Every term in the resultant sum will be multiplied by at least one of these derivatives. The only terms which singularities will be $\log(P_{ab})$, $P_{ab}^{-1}$ and $P_{ab}^{-2}$, and as each $P_{ab}$ is bounded away from 0 for large $t$, under the assumption that $\pi_i > 0$ for all $i$, these converge to finite limits. Thus, each term in the resultant sum will converge to 0 as required.
\subsubsection{Behaviour of \textit{S} near \textit{t = 0}}

\begin{lemma}
\label{lem:Better0}
    Define
\begin{equation}
    K = \sum_{a \neq b}\pi_a Q_{ab}
\end{equation}
Consider taking $Kt \to 0$ while keeping the stationary distribution $\pi$ and the ratios $\frac{Q_{ab}}{Q_{cd}}$ constant (i.e. one can change $t$ or scale the matrix $Q$ by a constant multiple). Then,
\begin{equation}
    S(t) \sim K t\log(Kt) +\mathcal{O}(Kt)  \quad \text{as $Kt \to 0$}
\end{equation}
\begin{equation}
    S'(t) \sim K \log(Kt) +o(K \log(Kt))  \quad \text{as $Kt \to 0$}
\end{equation}
and
\begin{equation}
    S''(t) \sim \frac{K}{t} + o(\frac{K}{t}) \quad \text{as $Kt \to 0$}
\end{equation}
\end{lemma}
\textbf{Proof:} All terms in sum defining $K$ have the same sign, and so $Kt \to 0$ implies that $K \pi_a Q_{ab} \to 0$ for all $a \neq b$. As $\pi$ is constant, this means that $K Q_{ab} \to 0$ for all $a \neq b$ and hence, summing these shows that $K Q_{aa} \to 0$ for all $a$.

The forward equations for a CTMC state that
\begin{equation}
    P'(t) = PQ \quad \text{and} \quad P(0) = I
\end{equation}
Thus, 
\begin{equation}
    P_{ab}(t) \sim \left\{\begin{matrix} Q_{ab}t + \mathcal{O}(Q_{ab}^2t^2) & \text{if $a \neq b$}\\
    (1-Q_{aa}t) + \mathcal{O}(Q_{aa}^2t^2) & \text{if $a=b$}\end{matrix}\right. \quad \text{as $Kt \to 0$}
    \end{equation}
Now, as $Kt \to 0$
\begin{align}
    S(t)&= \sum_{a,b} \pi_aP_{ab}(t)\log(P_{ab}(t))\\
    &\sim \sum_{a\neq b} \pi_a(Q_{ab}t + \mathcal{O}(Q_{ab}^2t^2))(\log(Q_{ab}t + \mathcal{O}(Q_{ab}^2 t^2))) + \sum_a \pi_a(1-Q_{ab}t)\log(1-Q_{ab}t)\\
    &\sim \sum_{a\neq b}\pi_a(Q_{ab}t + \mathcal{O}(K^2t^2))(\log(Q_{ab}t + \mathcal{O}(K^2 t^2))) + \mathcal{O}(Kt)\\
    &\sim  \sum_{a\neq b}\pi_aQ_{ab}t \log(Q_{ab}t) + \mathcal{O}(Kt)
\end{align}
Now, note that the value of $\frac{\pi_a Q_{ab}}{K}$ is constant (and hence $\mathcal{O}(1)$) by assumption. Thus,
\begin{equation}
    S(t)\sim \sum_{a \neq b}\pi_aQ_{ab}t \log(Kt) + \mathcal{O}(Kt) = Kt\log(Kt) + \mathcal{O}(Kt)
\end{equation}
which is the required result.

The derivatives follow by noting that the ignored terms are all of the form $(Kt)^n$ and $(Kt)^m\log(Kt)$ for $n \geq 1$ and $m \geq 2$. As these terms all have larger powers of $Kt$ and $Kt << 1$, the derivatives of these terms will be much smaller than the derivative of the leading order term, and hence we can simply differentiate the leading order term to find the leading order derivatives.

\end{document}